\documentclass[reprint,superscriptaddress,amsmath,amssymb,aps,prb]{revtex4-2}

\usepackage{graphicx}
\usepackage{dcolumn}
\usepackage{bm}
\usepackage{braket}
\usepackage{xcolor}
\usepackage{hyperref} 
\hypersetup{colorlinks, allcolors=blue}

\newcommand{\angstrom}{\text{\normalfont\AA}}

\begin{document}

\title{Topological properties of finite-size heterostructures of magnetic topological insulators and superconductors}

\author{Julian Legendre}
\affiliation{Department of Physics and Materials Science, University of Luxembourg, 1511 Luxembourg, Luxembourg}

\author{Eduárd Zsurka}
\affiliation{Department of Physics and Materials Science, University of Luxembourg, 1511 Luxembourg, Luxembourg}
\affiliation{Peter Grünberg Institute (PGI-9), Forschungszentrum Jülich, 52425 Jülich, Germany}
\affiliation{JARA-Fundamentals of Future Information Technology, Jülich-Aachen Research Alliance, Forschungszentrum Jülich and RWTH Aachen University, Germany}

\author{Daniele Di Miceli}
\affiliation{Institute for Cross-Disciplinary Physics and Complex Systems IFISC (CSIC-UIB), E-07122 Palma, Spain}
\affiliation{Department of Physics and Materials Science, University of Luxembourg, 1511 Luxembourg, Luxembourg}

\author{Llorenç Serra}
\affiliation{Institute for Cross-Disciplinary Physics and Complex Systems IFISC (CSIC-UIB), E-07122 Palma, Spain}
\affiliation{Department of Physics, University of the Balearic Islands, E-07122 Palma, Spain}

\author{Kristof Moors}
\affiliation{Peter Grünberg Institute (PGI-9), Forschungszentrum Jülich, 52425 Jülich, Germany}
\affiliation{JARA-Fundamentals of Future Information Technology, Jülich-Aachen Research Alliance, Forschungszentrum Jülich and RWTH Aachen University, Germany}

\author{Thomas L. Schmidt}
\affiliation{Department of Physics and Materials Science, University of Luxembourg, 1511 Luxembourg, Luxembourg}

\begin{abstract}
Heterostructures of magnetic topological insulators (MTIs) and superconductors (SCs) in two-dimensional (2D) slab and one-dimensional (1D) nanoribbon geometries have been predicted to host, respectively, chiral Majorana edge states (CMESs) and Majorana bound states (MBSs). We study the topological properties of such MTI/SC heterostructures upon variation of the geometry from wide slabs to quasi-1D nanoribbon systems and as a function of the chemical potential, the magnetic doping, and the induced superconducting pairing potential. To do so, we construct effective symmetry-constrained low-energy Hamiltonians accounting for the real-space confinement. For a nanoribbon geometry with finite width and length, we observe different phases characterized by CMESs, MBSs, as well as coexisting CMESs and MBSs, as the chemical potential, the magnetic doping and/or the width are varied.
\end{abstract}

\maketitle

\section{Introduction}

Topological superconductors are fascinating phases of matter which have stirred significant interest in the scientific community \cite{HasanKane10,QiZhang11,SatoAndo17}. These phases display gapped bulk states with superconducting pairing as well as topologically protected gapless surface states, which have been predicted to be Majorana states. The search for these quasi-particles has stimulated an intense research activity, primarily owing to their potential for quantum computing \cite{Kitaev01,Alicea12}. Nevertheless, proposed realizations of topological superconductors with large enough bulk gaps remain rare and Majorana states remain elusive and controversial.

Bringing together superconducting pairing and spin-orbit (SO) or SO-like interactions is a promising avenue for creating topological superconductivity. It has for instance been studied in superconductors with strong SO interactions \cite{Sato09,ZhangYaji18,Machida19}, in topological materials where doping with Nb, Sr or Cu yields a superconducting gap in the bulk \cite{HorWilliams10,Du17}, or in heterostructures combining strong SO interactions or SO-like interactions induced by a magnetic texture with a conventional superconductor \cite{FuKane08,FuKane09,LutchynSau10,OregRefael10,WangLiu12,Klinovaja13,NadjPerge14,WangZhou15,YangStano16,Mingyang18,ZengYongxin18}. Breaking time-reversal symmetry (TRS) is also often necessary for the emergence of low-dimensional surface states such as chiral edge states or bound states. 

%%%%%%%%%%%%%%%%%%%%%%%%%%%%%%%%%%%%%%%%%%%%%%%%%%%%%
\begin{figure*}
    \centering
    \includegraphics[width=\textwidth]{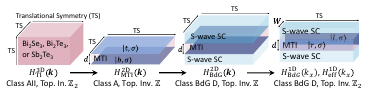}
    \caption{Schematic of the systems under consideration and the corresponding notations used in the main text. From left to right: a 3D TI with translation symmetry (TS) in the three directions of space; an MTI slab with TS only in the plane; an MTI slab with in-plane TS in contact with s-wave superconductors at its top and bottom surfaces; and an MTI slab with TS only along one direction of the plane, in contact with s-wave superconductors at its top and bottom surfaces. The notations we use for the Hamiltonian describing each system, along with the symmetry class of the Hamiltonian, appear respectively below each system schematic.}
    \label{fig:schema}
\end{figure*}
%%%%%%%%%%%%%%%%%%%%%%%%%%%%%%%%%%%%%%%%%%%%%%%%%%%%

TRS can also be broken by an external magnetic field, but this may not be compatible with superconductivity. It is thus desirable to explore intrinsic magnetism (e.g. via magnetic doping) of heterostructures as an alternative \cite{SatoNagahama24,UdayLippertz23,YiChan24,YuanYan24}. In our work, we study MTI/SC heterostructures where the interplay between superconducting pairing, spin-orbit coupling, and TRS breaking leads to the appearance of topological superconducting phases. Our study will apply to heterostructures consisting of $s$-wave SCs and magnetically doped compounds of the Bi$_2$Se$_3$ family. The effective realization of such heterostructures has recently shown promising progress \cite{SatoNagahama24,UdayLippertz23}. We will consider thin MTI films, which have become experimentally realizable over the past decade \cite{Landolt14, ZhangHe10, Neupane14}.
A comprehensive description of MTI thin films is achieved by the construction of a symmetry-constrained $\bf{k} \cdot \bf{p}$ Hamiltonian \cite{ZhangLiu09,QiZhang11}. Here, we introduce this model as a basis for the subsequent finite-size calculations and to relate our study to concrete material systems. We discuss its applicability for the system we investigate and we comment on the limits of such a model \cite{Forster15,Forster16,LiuYang23}.

In recent works, planar translation-invariant geometries and confined quasi-1D geometries of MTI/SC heterostructures have been suggested as a potential platform for Majorana physics. Such systems have been studied for specific values of the chemical potential $\mu$, of the strength $\lambda$ of the magnetic exchange interaction with the magnetic dopants, and for specific sizes \cite{ChenXie18,WangZhou15,ZengYongxin18}. Here, we provide a more comprehensive treatment, studying a wider region of the phase diagram in $(\mu,\lambda)$ space and the nature of the topological states when the in-plane size of the heterostructure is varied from infinitely large to finite. 
Moreover, we discuss the transition from chiral Majorana edge states (CMES) to Majorana bound states (MBS) and their respective localization. In the following, we will refer to planar translation-invariant geometries as ``slab geometries" and to in-plane confined geometries as ``nanoribbons". For nanoribbons with intermediate width, and depending on the magnitude of the magnetic exchange term in the MTI, we observe regions in the phase diagram where the low-energy sector hosts coexisting CMESs and MBSs. Our understanding of the topological properties associated with these finite-size systems, from two-dimensional to quasi-one-dimensional, is based on symmetry-constrained analytical low-energy models that we construct throughout this paper. These characterize the appearance of gapless surface (edge or end) states through the bulk-edge correspondence. 

This paper is organized as follows: In Sec.~\ref{sec2}, we introduce the effective models we use for the description of the MTI. First, we review the construction of a symmetry-constrained $\bf{k} \cdot \bf{p}$ Hamiltonian which characterizes a 3-dimensional (3D) topological insulator (TI). Then we construct effective models for a thin MTI slab system and for a nanoribbon geometry by considering the low-energy states arising from quantum confinement in the 3D model. We study the occurence of chiral edge modes as a function of the magnetic exchange term and as a function of the width of the nanoribbon.
In Sec.~\ref{MTI/SC}, we study the topological properties of an MTI/SC slab and then we investigate the occurrence of CMES and MBS in MTI/SC with nanoribbon geometries.

\section{Effective model for the MTI} \label{sec2}

\subsection{Symmetry-constrained \texorpdfstring{$\bf{k} \cdot \bf{p}$}{k dot p} Hamiltonian}

\begin{figure}[t]
    \includegraphics[width=\linewidth]{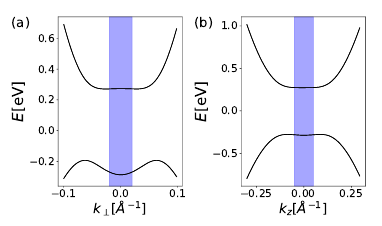}
    \caption{Low-energy spectrum of the 3D TI Hamiltonian~\eqref{ljehwbcrv}. (a) Spectrum as a function of the in-plane momentum $k_\perp \in \{k_x, k_y\}$ for $k_z=0$. (b) Spectrum as a function of $k_z$ for $k_\perp=0$. We use the parameters for Bi$_2$Se$_3$ given in Ref.~\cite{ZhangLiu09}: $C=-0.0068$ eV, $M=0.28$ eV, $A_1=2.2$ eV$\angstrom$, $A_2=4.1$ eV$\angstrom$, $B_1=10$ eV$\angstrom^2$, $B_2=56.6$ eV$\angstrom^2$, $D_1=1.3$ eV$\angstrom^2$, $D_2=19.6$ eV$\angstrom^2$. The blue shaded area indicates the momentum space region where the band structure of the Hamiltonian in the Eq.~\eqref{ljehwbcrv} is in good agreement with ab-initio calculations \cite{LiuQi10}.}
    \label{fig:Energies3D}
\end{figure}

A convenient way of describing the topological properties of a (3D) compound of the $\textrm{Bi}_2 \textrm{Se}_3$ family is via the following Hamiltonian \cite{Kane66,Winkler,ZhangLiu09,QiZhang11},
\begin{align} \label{ljehwbcrv}
H_{\textrm{TI}}^{3\textrm{D}}({\bf k})
&= H_0(k_z) + H_1(k_x,k_y), \\
H_0(k_z) \notag 
&= D_1 k_z^2  \mathbb{I} + \left(M-B_1 k_z^2 \right) \tau_z + A_1 k_z \tau_x \sigma_z, \\ \notag 
H_1(k_x,k_y)
&= \left( C +D_2k^2\right)  \mathbb{I} -B_2k^2 \tau_z \\ \notag 
&+ A_2 \tau_x \left( k_x \sigma_x + k_y \sigma_y \right),\notag
\end{align}
where $k^2 \equiv k_x^2+k_y^2$, $k_\pm \equiv k_x\pm i k_y$, and $C$, $M$, $D_1$, $D_2$, $B_1$, $B_2$, $A_1$, $A_2$, are real coefficients. This Hamiltonian acts in the (atomic) low-energy basis of states $\{\ket{+,\uparrow},\ket{-,\uparrow},\ket{+,\downarrow},\ket{-,\downarrow}\}$ through the Pauli matrices $\sigma_{x,y,z}$ and $\tau_{x,y,z}$ which act, respectively, in the spin $\{\uparrow,\downarrow\}$ and parity $\{+,-\}$ spaces. The parity operation, equivalent to inversion symmetry in 3D, acts on $\{\ket{+,\uparrow},\ket{-,\uparrow},\ket{+,\downarrow},\ket{-,\downarrow}\}$ according to the matrix representation $ \tau^z$ and flips the momentum vector, leaving the Hamiltonian invariant: $\tau^z H_{\textrm{TI}}^{3\textrm{D}}(- {\bf k}) \tau^z = H_{\textrm{TI}}^{3\textrm{D}}({\bf k})$.
This Hamiltonian is valid in the vicinity of the $\Gamma$ point, where it describes the low-energy properties of the system. In Fig.~\ref{fig:Energies3D}, we show the eigenenergies of this Hamiltonian.

This Hamiltonian relies on a long-wavelength approximation and is \textit{a priori} valid only for finite systems above a certain size. The coefficients appearing in the Hamiltonian can be determined by fitting the energy spectrum to experimental data or extracting the parameters from \textit{ab-initio} calculations (see Ref.~\cite{QiZhang11}). The distance from the $\Gamma$ point up to which this model Hamiltonian is a good approximation allows an estimate of the possible size of the system.
The band structure of the Hamiltonian (\ref{ljehwbcrv}) is in good agreement with \textit{ab-initio} calculations for $|k_x| \lesssim 0.02$ $\angstrom {}^{-1}$, $|k_y| \lesssim 0.02$ $\angstrom{}^{-1}$, and $|k_z| \lesssim 0.05 \angstrom{}^{-1}$ \cite{LiuLiang10}. This implies that the system thickness $d$ in the $z$ direction should be larger than $20$ $\angstrom$ while the width $W$ and the length $L$, respectively, in the $y$ direction and $x$-directions should be larger than $50$ $\angstrom$.

It is also important to note that $H_{\textrm{TI}}^{3\textrm{D}}({\bf k})$ is constrained by the symmetries of the system, specifically TRS, inversion symmetry, and the three-fold rotation symmetry around the $z$ axis. The atomic states used to construct the Hamiltonian are based on quintuple layer unit cells which respect the aforementioned symmetries~\cite{QiZhang11}. For the materials we consider, the thickness (along the $z$ direction) of a quintuple layer is approximately $1$ nm. In the following, we will consider systems with widths $W$ and lengths $L$ larger than $40$ nm, which is much larger than the length of a unit cell in the $x$ and $y$-directions (below $1$ nm). This allows us to study a continuum of lengths above $40$ nm along these directions. We notice that $W \sim 40$ nm or $L \sim 40$ nm is a reasonable lower limit of what is currently achieved experimentally. We will consider slabs and nanoribbons of $\textrm{Bi}_2 \textrm{Se}_3$ family compounds consisting of two or more quintuple layers (thickness $d$ larger than $20$ $\angstrom$).

%%%%%%%%%%%%%%%%%%%%%%%%%%%%%%%%%%%%%%%%%%%%%%%%%%%%%
%%%%%%%%%%%%%%%%%%%%%%%%%%%%%%%%%%%%%%%%%%%%%%%%%
%%%%%%%%%%%%%%%%%%%%%%%%%%%%%%%%%%%%%%%%%%%%%%%%%%%%%

\subsection{Slab geometry} \label{slabjcdenv}
\begin{figure}[t]
    \includegraphics[width=\linewidth]{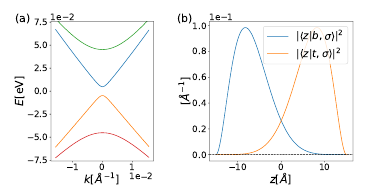}
    \caption{(a) Energy spectrum of the translation-invariant MTI slab model in the QAH regime for thickness $d=30$ $\angstrom$, $\lambda = 25$ meV and $\mu=0$. (b) Probability distributions $|\braket{z|l,\sigma}|^2$, where $l\in \{t,b\}$, associated with the top and bottom surface states of the slab for a thickness $d=30$ $\angstrom$.}
    \label{fig:E_lambda}
\end{figure}

Let us consider a TI slab (a planar translation-invariant geometry) with a finite thickness $d$ along the $z$ direction, with bottom and top surfaces of the slab located, respectively, at $z=0$ and $z=d$. We replace $k_z \to -i \partial_z$, impose vanishing wave functions as boundary conditions at $z= 0$ and $z= d$, and denote by $E_+$ and $E_-$ the two lowest eigenenergies of $H_0(-i\partial_z)$, each of which is twofold degenerate because of TRS. Magnetic doping can be accounted for by adding a TRS-breaking Zeeman term $\lambda \sigma_z$; the system is now an MTI. The respective eigenstates of the Hamiltonian will be denoted by $\ket{\varphi^\sigma}$ and $\ket{\chi^\sigma}$, where $\sigma \in \{\uparrow,\downarrow\}$.
Projecting $H_{\textrm{TI}}^{3\textrm{D}}({\bf k})$ on the basis $\left\{ \ket{\varphi^\uparrow}, \ket{\chi^\uparrow}, \ket{\varphi^\downarrow}, \ket{\chi^\downarrow}\right\}$ gives rise to a $4 \times 4$ Hamiltonian which describes an MTI slab \cite{ZhouLu08,LuShan10} (we use $\hbar = 1$ in the following),
\begin{align} \label{eq:hyedfpobd}
    H_{\textrm{MTI}}^{2\textrm{D}}({\bf k}) &=
\begin{pmatrix}
h_+({\bf k})  & 0 \\
0 & h_-^*({\bf k})
\end{pmatrix}, \\
h_\pm({\bf k}) &\equiv - \mu - D k^2 + v_\mathrm{F} \left( k_y \sigma_x -  k_x \sigma_y \right) \notag \\
&+  \left(m_0 \pm \lambda +m_1 k^2 \right)\sigma_z, \notag 
\end{align}
where the coefficients are given by
\begin{align} \label{eq:coefsha}
    -\mu &\equiv \left(E_- + E_+\right)/2 + C, \notag \\
    D &\equiv \left(B_2/2\right) \left( \bra{\varphi^\uparrow} \tau_z \ket{\varphi^\uparrow} + \bra{\chi^\uparrow} \tau_z \ket{\chi^\uparrow} \right) - D_2, \notag \\
    v_\mathrm{F} &\equiv -i A_2 \bra{\varphi^\uparrow} \tau_x \ket{\chi^\downarrow} \in \mathbb{R}, \notag \\
    2 m_0 &\equiv E_- - E_+, \notag \\
    m_1 &\equiv- \left(B_2/2\right) \left( \bra{\varphi^\uparrow} \tau_z \ket{\varphi^\uparrow} - \bra{\chi^\uparrow} \tau_z \ket{\chi^\uparrow} \right).
\end{align}
In Table~\ref{tab1} we list the values of the coefficients $m_0$, $m_1$, $D$ and $v_\mathrm{F}$ for several values of the thickness, starting from two quintuple layers, and calculated from Eq.~\eqref{eq:coefsha}, using the parameters of the 3D ${\bf k} \cdot {\bf p}$ Hamiltonian~\eqref{ljehwbcrv} for bulk Bi$_2$Se$_3$~\cite{ZhangLiu09}. A diagonalization of $H_{\textrm{MTI}}^{2\textrm{D}}({\bf k})$ yields four energy bands with a finite gap at the $\Gamma$ point. This is shown in Fig.~\ref{fig:E_lambda}(a) where, for simplicity, we considered the chemical potential of the system to be tuned to $\mu =0$.

%%%%%%%%%%%%%%%%%%%%%%%%%%%%%%%%%%%%%%%%%%%%%%%%%%%%%%%%%%%%%%%%%%%%%%
\begin{table}[b]
\caption{\label{tab1}Values of the coefficients $m_0$, $m_1$, $D$ and $v_\mathrm{F}$ for a slab with thickness between two and seven quintuple layers. Above seven quintuple layers, $m_0$ and $m_1$ have negligible magnitudes and $D$ and $v_\mathrm{F}$ become approximately constant.}
\begin{ruledtabular}
\begin{tabular}{cccccc}
d[\angstrom]& $m_0$[eV] & $m_1$[eV$\angstrom^2]$ & $D$[eV$\angstrom^2]$ & $ v_\mathrm{F}$[eV$\angstrom]$ \\
\colrule
$20 $ & $6.9 \times 10^{-2}$ 
&  $45.48$ & $-17.35$ & 4.09\\
$30 $  &  $-2.0 \times 10^{-2}$ & $19.81$ & $-12.64$ & 4.06\\
$40$ &  $-1.1 \times 10^{-2}$  &  $-2.82$ & $-12.05$  & 4.06\\
$50$ &  $-7.5 \times 10^{-5}$  &  $-4.29$ & $-12.29$  & 4.06\\
$60$ &  $1.2 \times 10^{-3}$  &  $-0.59$ & $-12.24$  & 4.06\\
$70$ &  $2.6 \times 10^{-4}$  &  $0.51$ & $-12.24$   & 4.06\\
\end{tabular}
\end{ruledtabular}
\end{table}
%%%%%%%%%%%%%%%%%%%%%%%%%%%%%%%%%%%%%%%%%%%%%%%%%%%%%%%%%%%%%%%%%%%%

The eigenstates $\ket{\varphi^\sigma}$ and $\ket{\chi^\sigma}$ allow the construction of states $\ket{t,\sigma}$ and $\ket{b,\sigma}$ localized near the top and bottom surfaces, respectively. These are indeed given by $\ket{t,\sigma} = \left(\ket{\varphi^\sigma} + \ket{\chi^\sigma}\right)/\sqrt{2}$ and $\ket{b,\sigma} = \left(\ket{\varphi^\sigma} - \ket{\chi^\sigma}\right)/\sqrt{2}$ and have localization lengths $\sim 10$ $\angstrom$ (see Fig.~\ref{fig:E_lambda}(b)).
In this basis of states, the Hamiltonian reads
\begin{align} \label{eq:tb}
\begin{split}
\tilde{H}_{\textrm{MTI}}^{2\textrm{D}}({\bf k}) = & -\mu - D k^2 +  v_\mathrm{F} \left( k_y \sigma_x -  k_x \sigma_y \right) \tilde{\tau}_z \\ &+ \lambda \sigma_z + m({\bf k}) \tilde{\tau}_x,
\end{split}
\end{align}
where $m({\bf k}) = m_0 +m_1 k^2 $ and the Pauli matrices $\tilde{\tau}_{x,y,z}$ and $\sigma_{x,y,z}$ act, respectively, on the top/bottom ($t/s)$ degree of freedom and the spin degree of freedom. Eq.~\eqref{eq:tb} is the starting point of Sec.~\ref{MTI/SC}, where we study an MTI/SC heterostructure.

In case of magnetic doping ($\lambda \neq 0$), the symmetry class of the Hamiltonian is denoted A (unitary symmetry class) and the topological phase is characterized by an integer Chern number \cite{Schnyder08}. In contrast, at $\lambda = 0$, the Hamiltonian has TRS and fits in the symmetry class AII (symplectic symmetry class) where the topological sector is characterized by a $\mathbb{Z}_2$ topological invariant.

Determining the value of the topological invariant requires not only the low-energy eigenstates but also information at large momenta. However, determining the difference between the topological invariants characterizing two phases separated by the closing of the gap is possible: in this case, we only need the eigenstates around the $\Gamma$ point, where the gap closes, which we determine from the Hamiltonian in Eq.~\eqref{eq:hyedfpobd}. These eigenstates and thus the topological invariants do not depend on $\mu$ or $D$ since those enter the Hamiltonian with an identity matrix. Hence, let us consider the case $\mu =0$.

The low-energy model can predict the occurrence of topological phase transitions via a sign change of the effective gap $m_0 \pm \lambda$. 
At $\lambda = 0$, the sign change of the gap $m_0$, when increasing the thickness from 2 to 3 quintuple layers, shows a transition between a trivial insulating and a quantum spin Hall insulating phase (see Table~\ref{tab1}). At $0 < |\lambda| < |m_0|$, the Chern number vanishes, indicating a trivial phase. In contrast, a sufficiently large magnetic exchange term $|\lambda|>|m_0|$ yields a quantum anomalous Hall (QAH) phase, independently of the sign and magnitude of $m_0$, as was also observed in Ref.~\cite{WangLian13}.

%%%%%%%%%%%%%%%%%%%%%%%%%%%%%%%%%%%%%%%%%%%%%%%%%%%%%
%%%%%%%%%%%%%%%%%%%%%%%%%%%%%%%%%%%%%%%%%%%%%%%%%%%%

Whether or not the materials we consider indeed display a quantum spin Hall phase (without doping) below 6 quintuple layers is still under debate in the literature. For instance, for Bi$_2$Se$_3$ it has been argued that the Coulomb interaction leads to a significant hybridization of the surface states, which could open a trivial gap below 6 quintuple layers \cite{WangZhou19,LiuYang23}. GW computations performed for Bi$_2$Se$_3$ also concluded that the gap below 6 quintuple layers is trivial, but a gap inversion remains for Bi$_2$Te$_3$ \cite{Forster15}. Our following study will apply to a thin slab in (or near) the QAH regime, which is attainable independently of the sign of $m_0$ for a large enough magnetic exchange magnitude $\lambda$. Experimental studies on such systems have already been performed, and the QAH phase has been characterized, e.g., in Cr- and V-doped (Bi,Sb)$_2$Te$_3$ \cite{ChangZhang2013}.

\subsection{One-dimensional model for an MTI nanoribbon} \label{sec:hjsdfco}

Next, we consider a 2D nanoribbon geometry with a finite width $W$ along the $y$ direction with left and right edges of the slab respectively located at $y=0$ and $y=W$. Moreover, we assume $\lambda>|m_0|$ which would correspond to the QAH regime for $W \rightarrow \infty$. The goal of this section is to estimate the critical width above which the nanoribbon effectively enters a QAH phase, manifested in one chiral state at each edge of the slab. Below this width, the hybridization of edge states on opposite edges becomes significant.

\begin{figure}[t]
    \includegraphics[width=\linewidth]{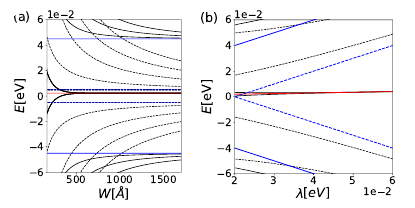}
    \caption{Low-energy spectrum of the MTI nanoribbon at the $\Gamma$ point ($k_x=0$) with $d = 30$ $\angstrom$ (a) for $\lambda=25$ meV as a function of the width $W$ and
    (b) for the width $W=600$ $\angstrom$ as a function of the magnetic exchange magnitude $\lambda$. We used the coefficients $m_0$, $m_1$, and $v_\mathrm{F}$ which appear in Table~\ref{tab1}, we considered $D=-1 \text{eV}\angstrom^2$, and the energy shift $-\mu$ appearing in the Eq.~\eqref{eq:hyedfpobd} is discarded for simplicity. The solid and the dashed black lines are the energies associated to $ h_-^*(k_x,-i \partial_y)$ and $h_+(k_x,-i \partial_y)$, respectively. The solid and the dashed blue lines are the bulk energies $\pm \left(m_0 - \lambda\right)$ and $\pm \left(m_0 + \lambda\right)$, respectively. The red line is the energy $(m_0-\lambda) D/m_1$.}
    \label{fig:lowE_d}
\end{figure}

The Hamiltonian~\eqref{eq:hyedfpobd} describes the 2D MTI slab and consists of the two independent blocks $h_+({\bf k})$ and $ h_-^*({\bf k})$. For the description of the nanoribbon, we substitute $k_y \to -i \partial_y$ and we impose vanishing wave functions at the edges of the slab. In the following, we assume $\left(m_0 - \lambda \right) m_1 <0$ without loss of generality, since the case $\left(m_0 - \lambda \right) m_1 >0$ would only exchange the roles of $ h_-^*(k_x,-i \partial_y)$ and $ h_+(k_x,-i \partial_y)$. The block $ h_-^*(k_x,-i \partial_y)$ has an inverted mass gap $\left(m_0 - \lambda \right) m_1<0$, which results in a pair of edge states arising from the quantum confinement along the $y$ direction. The associated energies at $k_x=0$ converge to $-\mu+\left(m_0-\lambda\right)D/m_1$ when $W$ or $\lambda$ is increased as shown respectively in Fig.~\ref{fig:lowE_d}(a) or Fig.~\ref{fig:lowE_d}(b) (red line). 
In contrast, $h_+(k_x,-i \partial_y)$ has a normal mass gap $\left(m_0 + \lambda \right) m_1>0$, so for the parameter regime we consider its low-energy states arising from the quantum confinement are topological trivial. They have energies greater than $\left|m_0 + \lambda  \right|$ or smaller than $-\left|m_0 + \lambda \right|$, respectively, as shown in Fig.~\ref{fig:lowE_d}. 

For non-vanishing $D$, the energies of both topological edge states are shifted compared to the energies of the bulk states. As a consequence, for large enough $D$, i.e., $|\left(m_0-\lambda\right)D/m_1|>\left|m_0 + \lambda  \right|$, the low-energy states at the $\Gamma$ point are outside the energy gap. As we would like to build a simple ribbon model, we will consider small values of $D$ in the following, such that the low-energy states at the $\Gamma$ point are in the gap. Hence, in Fig.~\ref{fig:lowE_d} we have assumed $D=-1 \text{eV}\angstrom$. For a description of the effects of larger values of $D$, we refer to Ref.~\cite{zsurka24}.

At $\mu= \left(m_0-\lambda\right)D/m_1$, where the Fermi energy corresponds to the red line in Fig.~\ref{fig:lowE_d}, we can construct a Hamiltonian which describes the edge states around the Fermi level by only considering the block $ h_-^*(k_x,-i \partial_y)$, as it describes the relevant low-energy physics. Substituting $k_y \to -i \partial_y$ and imposing vanishing wave functions at the edges of the slab, $h_-^*$ becomes,
\begin{align} \label{eq:8}
    &h_-^* = h_{0-}(-i \partial_y) + h_{1-}(k_x), 
\\
    &h_{0-}(-i \partial_y) = D \partial_y^2 -i  v_\mathrm{F} \sigma_x \partial_y  +  \left(m_0 - \lambda -m_1 \partial_y^2 \right)\sigma_z,\notag 
\\
    &h_{1-}(k_x) = -\mu - D k_x^2 +  v_\mathrm{F}  k_x \sigma_y   +m_1 k_x^2 \sigma_z.\notag 
\end{align}
Let us denote the lowest eigenergies at $k_x=0$ by $E_+^y$ and $E_-^y$, and the associated eigenstates by $\chi_y$ and $\varphi_y$. Similarly to the previous section, we project $ h_{1-}(k_x)$ on the low-energy eigenstates of $h_{0-}(-i \partial_y)$ and thus obtain an effective 1D Hamiltonian describing the slab with finite width $W$ at low energies (see App.~\ref{app:LowEProj})
\begin{align} \label{eq:HMTIprojectedW}
h_{-}^{1\textrm{D}}(k_x) =&-\mu + E_1 -\tilde{D} k_x^2  -  \tilde{v}_\mathrm{F} k_x \tilde{\sigma}_x + \left(\tilde{m}_0 + \tilde{m}_1 k_x^2 \right) \tilde{\sigma}_z,
\end{align}
where $\tilde{\sigma}_{x,z}$ are Pauli matrices acting in the $\left\{\varphi_y,\chi_y\right\}$ basis and the parameters are given by
\begin{align}
    2 E_1 &= E_+^y + E_-^y, \notag \\
    2\tilde{m}_0 &\equiv-E_+^y + E_-^y \notag \\
    2\tilde{D} &\equiv 2D-m_1\left(\bra{\varphi_y} \sigma_z \ket{\varphi_y} +\bra{\chi_y} \sigma_z \ket{\chi_y} \right), \notag\\
    2\tilde{m}_1 &\equiv m_1 \left(\bra{\varphi_y} \sigma_z \ket{\varphi_y} -\bra{\chi_y} \sigma_z \ket{\chi_y} \right), \notag \\
    \tilde{v}_\mathrm{F} &\equiv v_\mathrm{F} \bra{\varphi_y} \sigma_y \ket{\chi_y}.    
\end{align}
%%%%%%%%%%%%%%%%%%%%%%%%%%%%%%%%%%%%%%%%%%%%%%%%%%%%%
%%%%%%%%%%%%%%%%%%%%%%%%%%%%%%%%%%%%%%%%%%%%%%%%%

\begin{figure}[t]
  \includegraphics[width=\linewidth]{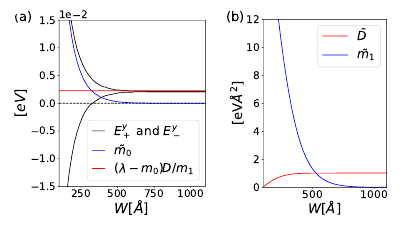}

    \caption{Effects of the finite MTI slab width (a) on the low eigenenergies $E_{\pm}^y$ and the energy gap $\tilde{m}_0 = (E_+-E_-)/2$, (b) on the coefficients $\tilde{D}$, $\tilde{m}_1$, and on the edge Fermi velocity $\tilde{v}_\mathrm{F}$. The value of the magnetic exchange parameter is $\lambda=25$ meV thickness of the slab is $d = 30$ $\angstrom$ and the values of the coefficients $m_0$, $m_1$, and $v_\mathrm{F}$ are taken from Table~\ref{tab1} and we considered $D=-1 \text{eV}\angstrom^2$.}
    \label{fig:energies_over_w_d30}
\end{figure}

The coefficients $\tilde{D}$ and $\tilde{m}_1$ are plotted as functions of the width $W$ in Fig.~\ref{fig:energies_over_w_d30}. The coefficient $\tilde{v}_\mathrm{F}$ varies only very weakly: for the parameters considered in Fig.~\ref{fig:energies_over_w_d30}, $\tilde{v}_\mathrm{F} = 1 \text{eV}\angstrom$ and it has a maximum variation $\sim 10^{-3} \text{eV}\angstrom$ over the range of $W$ investigated.

Next, we introduce the states $\ket{L} = \left(\ket{\varphi_y} + \ket{\chi_y}\right)/\sqrt{2}$ and $\ket{R} = \left(\ket{\varphi_y} - \ket{\chi_y}\right)/\sqrt{2}$ which are localized on the left and right edges of the ribbon, respectively. The parameter $\tilde{m}_0$ represents the overlap between both edge states, so it determines the topological phase of the slab. For small overlap, in the QAH phase, one chiral state appears at each edge. For large $\tilde{m}_0$, the overlap of these edge states brings the system into a topologically trivial phase. In this 1D limit, a symmetry-class A Hamiltonian has indeed trivial topological properties \cite{Schnyder08}.

From Fig.~\ref{fig:energies_over_w_d30}, we see that for $W \lesssim 500$ $\angstrom$, $\tilde{m}_0$ cannot be neglected compared to the other terms in the Hamiltonian, so the 1D system becomes topologically trivial. For $W \gtrsim 500$ $\angstrom$, on the other hand, $\tilde{m}_0$ is negligible, so the slab enters a QAH phase. 

%%%%%%%%%%%%%%%%%%%%%%%%%%%%%%%%%%%%%%%%%%%%%%%%%%%%%
%%%%%%%%%%%%%%%%%%%%%%%%%%%%%%%%%%%%%%%%%%%%%%%%%%%%

\section{MTI/SC heterostructures} \label{MTI/SC}

\subsection{Surface states of the MTI and proximity effect}

We now consider the superconducting proximity effect \cite{Sitthison14,WangZhou15} originating from the contact with $s$-wave superconductors on the top and bottom surfaces of the MTI slab. We assume that they give rise to the pairing potentials $\Delta_1$ and $\Delta_2$, respectively, with two real parameters $\Delta_1$ and $\Delta_2$. This is described by extending the Hamiltonian~\eqref{eq:tb} to the Nambu basis,
\begin{align} \label{eq:ezrfgz} 
\begin{split}
    H_{\textrm{BdG}}^{2\textrm{D}}({\bf k}) 
&=  
     v_\mathrm{F} k_y \sigma_x \tilde{\tau}_z  + \bigg[ - v_\mathrm{F} k_x \sigma_y \tilde{\tau}_z + \lambda \sigma_z \\ 
&+ 
     m({\bf k})  \tilde{\tau}_x  - \mu - D k^2 \bigg] \gamma_z \\ 
&- 
    \left( \dfrac{\Delta_1 + \Delta_2}{2}  + \dfrac{\Delta_1 - \Delta_2}{2} \tilde{\tau}_z \right) \sigma_y \gamma_y.
\end{split}
\end{align}
where the Pauli matrices $\gamma_{y,z}$ act in the particle-hole space. Expressed in this form, it is straightforward to check that the Hamiltonian has particle-hole symmetry $\mathcal{P} H_{\textrm{BdG}}^{2\textrm{D}}({\bf k})  \mathcal{P}^{-1} = -H_{\textrm{BdG}}^{2\textrm{D}}(-{\bf k}) $ and at $\lambda=0$ it also has TRS $\Theta H_{\textrm{BdG}}^{2\textrm{D}}({\bf k})  \Theta^{-1} = H_{\textrm{BdG}}^{2\textrm{D}}(-{\bf k}) $, where $\Theta = i \sigma_y \mathcal{K}$, $\mathcal{P} = \gamma_x \mathcal{K}$, and $\mathcal{K}$ is the complex conjugation operator.

\subsection{Topological properties for the slab geometry} \label{MTISC_TransInv}

\begin{figure}[t]
    \includegraphics[width=\linewidth]{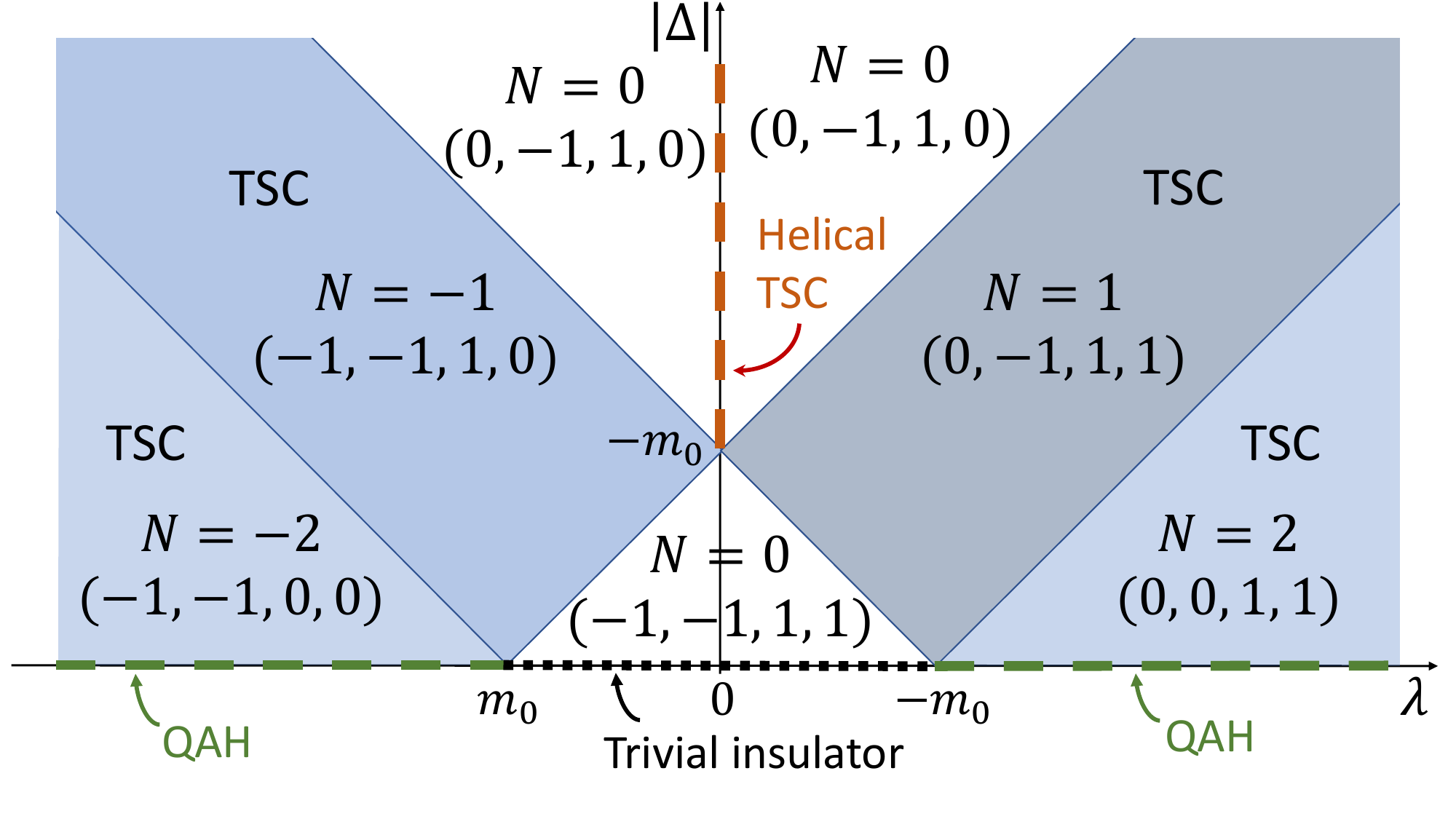}

    \caption{Topological phase diagram of the MTI/SC heterostructure at $m_0<0$, $m_1>0$, $\Delta_1=-\Delta_2$, $\mu=0$ and $D =0$. $N$ is the topological invariant and the four-components list given below N is $(N_{+}^+,N_{-}^+,N_{+}^-,N_{-}^- )$ (see details in Sec.~\ref{MTISC_TransInv}). The (helical) topological superconducting phases are indicated by ``(Helical) TSC" and the quantum anomalous Hall phases by ``QAH".}
    \label{fig41}
\end{figure}

For a slab geometry, the topological phase transitions are signaled by the change of the 2D bulk invariant of the model~\eqref{eq:ezrfgz} and happen at the gap closing points. For the special case $\Delta_1=-\Delta_2$, $\mu=0$ and $D =0$ the Hamiltonian can be simplified to (see App.~\ref{app:DiagBlocksH})
\begin{align}\label{eq:ham_decoupled}
    H_{\textrm{BdG}}^{2\textrm{D}}&= \sum_{{\bf k}} \sum_{\eta=\pm} \sum_{\kappa=\pm}  \psi_{\kappa,\bf k}^{\eta \dagger} h_\kappa^{\eta}( {\bf k}) \psi_{\kappa,\bf k}^{\eta} , \\
    h_\kappa^{\eta}({\bf k}) &=   v_\mathrm{F} \left(  k_y \sigma_x  - \kappa 
 k_x \sigma_y \right) +  m_\kappa^\eta({\bf k}) \sigma_z, \notag 
\end{align}
with $m_\kappa^\eta({\bf k}) = m_{0,\kappa}^\eta + m_{1,\kappa}^\eta k^2 $ and $m_{0,\kappa}^\eta = \kappa \lambda  + \eta \kappa m_0   + |\Delta| $, and $m_{1,\kappa}^\eta = \eta \kappa m_1$.
It is then straightforward to calculate the topological invariant \cite{WangZhou15}. The Hamiltonian is a sum of independent massive Dirac Hamiltonians, so the topological invariant is the sum $N=\sum_{\eta,\kappa} N_\kappa^\eta$ of the winding numbers $N_\kappa^\eta$ associated with each of the independent Hamiltonians $ h_\kappa^{\eta}({\bf k}) $. At fixed $m_1$, the cases $m_0>0$ and $m_0<0$ both result in the same phase diagram for $N$, even though the decompositions in terms of $N_\kappa^\eta$ are different. In Fig.~\ref{fig41}, we show this phase diagram along with the details of the decomposition in terms of $N_\kappa^\eta$ for $m_0<0$ and $m_1>0$. 

%%%%%%%%%%%%%%%%%%%%%%%%%%%%%%%%%%%%%%%%%%%%%%%%%%%%%
%%%%%%%%%%%%%%%%%%%%%%%%%%%%%%%%%%%%%%%%%%%%%%%%%%%%

Building on this, we can extend the phase diagram for $\Delta_1 \neq -\Delta_2$, $\mu \neq 0$ and $D \neq 0$ by considering the phase boundaries. As topological phase transitions happen only at the gap closing points, we find that they are described by the equation
\begin{align} \label{eq:iyhwedlof}
\begin{split}
    &\alpha^2 \Delta^4  + \Delta^2 \left[2 \alpha m_0^2- \left(1+ \alpha^2\right) \left(\lambda^2-\mu^2\right)\right] \\ &+ \left(\lambda^2+\mu^2 - m_0^2\right)^2-4\lambda^2 \mu^2 =0,
\end{split}
\end{align}
with $\alpha \equiv \Delta_2/\Delta_1$. This result generalizes the study performed in Ref.~\cite{WangZhou15} where an equation for the phase boundaries was given for $\mu=0$. 
The $D k^2 $ term, being proportional to $\gamma_z$ and involving only terms of second order in the momentum, does not influence the phase boundaries.

\subsection{Ribbon geometry for \texorpdfstring{$\mu=0$}{mu=0}, \texorpdfstring{$\Delta_1=-\Delta_2$}{Delta1=-Delta2}} \label{sec:hjsdfco1}

Next we investigate the effect of in-plane confinement on the topological properties of the MTI/SC heterostructure for the case where $\mu=0$ and $\Delta_1=-\Delta_2$. We show that for a confined geometry, the topological properties depend on the decomposition of $N$ in terms of $N_\kappa^\eta$, and on the mass values $m_{0,\kappa}^\eta$. This is in contrast to the translation-invariant geometry, where $N$ alone determines the topological properties at $\lambda \neq 0$.

\begin{figure}
    \includegraphics[width=\linewidth]{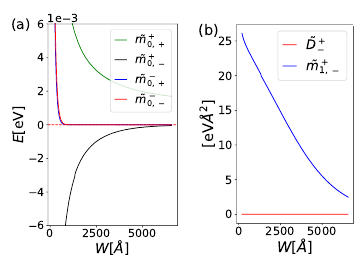}

    \caption{(a) Low-energy spectrum of an MTI/SC ribbon heterostructure as a function of the width $W$ for $\mu=0$ and $\Delta_1=-\Delta_2$. (b) The coefficients $\tilde{D}_{-}^{+}$ and $\tilde{m}_{1,-}^{+} $ are associated to the 1D bulk Hamiltonian which describes the physics of the low-energy states $\varphi_{y,-}^{+}$ and $\chi_{y,-}^{+}$. The thickness of the slab is $30$ $\angstrom$, the values of the coefficients $m_0$, $m_1$, $v_\mathrm{F}$ are taken from Table~\ref{tab1}, $D=0$, $\lambda=19$ meV and $|\Delta|=2$ meV.}
    \label{fig5}
\end{figure}

\begin{figure}
  \includegraphics[width=\linewidth]{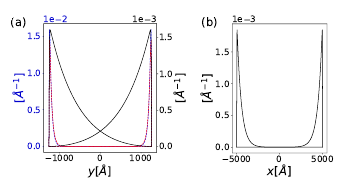}
    
    \caption{(a) Localization of the chiral edge states along the $y$ direction in an MTI/SC nanoribbon which is translation-invariant along the $x$ direction. The dashed lines represent $\ket{\varphi_{y,+}^{-}}$, $\ket{\chi_{y,+}^{-}}$, $\ket{\varphi_{y,-}^{-}}$, and $\ket{\chi_{y,-}^{-}}$ and the solid lines $\ket{\varphi_{y,-}^{+}}$, and $\ket{\chi_{y,-}^{+}}$. The left (blue) vertical axis refers to the dashed lines while the right (black) vertical axis refers to the solid black lines. (b) Localization of the end states $\ket{\varphi_{x,-}^{+}}$, and $\ket{\chi_{x,-}^{+}}$ along the $x$ direction in an MTI/SC nanoribbon of finite length $L=1 \mu$m. The thickness of the MTI slab is $d=30$ $\angstrom$ and its width is $W=2500$ $\angstrom$. The values of the coefficients $m_0$, $m_1$, $v_\mathrm{F}$ are taken from Table~\ref{tab1} and we used $D=0$, $\lambda=19$ meV and $|\Delta|=2$ meV.}
    \label{fig:DoS_simpleN1}
\end{figure}

First, we consider a ribbon geometry with width $W$ along the $y$ direction. The calculation performed in Sec.~\ref{sec:hjsdfco} can be easily adapted since the Hamiltonian~\eqref{eq:ham_decoupled} is a sum of independent massive Dirac Hamiltonians.
For each term where the mass $m_{0,\kappa}^\eta$ and the parameter $m_{1,\kappa}^\eta$ correspond to the inverted regime, i.e., $m_{0,\kappa}^\eta m_{1,\kappa}^\eta<0$, a pair of low-energy states appears, which we denote by $\ket{\varphi_{y,\kappa}^{\eta}}$ and $\ket{\chi_{y,\kappa}^{\eta}}$ and which have the respective energies $\Tilde{m}_{0,\kappa}^\eta $ and $- \Tilde{m}_{0,\kappa}^\eta $. These energies are plotted in Fig.~\ref{fig5} as a function of the width of the ribbon. Figure~\ref{fig:DoS_simpleN1}(a) shows the localization of $\ket{\varphi_{y,\kappa}^{\eta}}$ and $\ket{\chi_{y,\kappa}^{\eta}}$ along the $y$ direction in an MTI/SC nanoribbon with width $W=2500$ $\angstrom$. One finds that $\Tilde{m}_{0,\kappa}^\eta $ decreases when $m_{0,\kappa}^\eta$ increases and that the overlap between $\ket{\varphi_{y,\kappa}^{\eta}}$ and $\ket{\chi_{y,\kappa}^{\eta}}$ is proportional to $\Tilde{m}_{0,\kappa}^\eta $ (see Fig.~\ref{fig:DoS_simpleN1}(a) and Fig.~\ref{fig5}(a)). 

For large regions of the phase diagram the masses $m_{0,\kappa}^\eta$ can differ significantly. At intermediate widths $W$, this causes the coexistence of chiral low-energy states strongly localized at the edges with states which overlap along the $y$ direction \cite{QiHughes10}. For instance, in Fig.~\ref{fig5}, for $1000$ $\angstrom < W < 5000$ $\angstrom$, $\tilde{m}_{0,+}^{-}$ (blue line) and $\tilde{m}_{0,-}^{-}$ (red line) are small, so we expect that $\ket{\varphi_{y,+}^{-}}$, $\ket{\chi_{y,+}^{-}}$, $\ket{\varphi_{y,-}^{-}}$, and $\ket{\chi_{y,-}^{-}}$ are edge states. In contrast, $\tilde{m}_{0,-}^{+}$ (black line) is larger so the states $\ket{\varphi_{y,-}^{+}}$, and $\ket{\chi_{y,-}^{+}}$ overlap along $y$. 
The states overlaping along the $y$ direction are then described by a 1D bulk Hamiltonian, with parameters $\tilde{m}_{0,\kappa}^{\eta} $ and $ \tilde{m}_{1,\kappa}^{\eta} = m_{1,\kappa}^{\eta} (\bra{\varphi_{y,\kappa}^{\eta}} \sigma_z \ket{\varphi_{y,\kappa}^{\eta}} -\bra{\chi_{y,\kappa}^{\eta}} \sigma_z \ket{\chi_{y,\kappa}^{\eta}} )/2$ which depends on $W$.

Next, we also consider confinement along the $x$ direction such that the length in the $x$ direction satisfies $L \gg W$. In this case, we find that the overlapping states $\ket{\varphi_{y,\kappa}^{\eta}}$ and $\ket{\chi_{y,\kappa}^{\eta}}$ give rise to new low-energy states $\ket{\varphi_{x,\kappa}^{\eta}}$ and $\ket{\chi_{x,\kappa}^{\eta}}$ which are localized at both ends of the ribbon if $\tilde{m}_{0,\kappa}^\eta \tilde{m}_{1,\kappa}^\eta<0$ (see Fig.~\ref{fig:DoS_simpleN1}(b)). These states are Majorana bound states (MBS), which arise from the confinement of a 1D BdG D Hamiltonian with topologically non-trivial $\mathbb{Z}_2$ number.

We are interested only in states with negligible energy, i.e., with energy below a small energy threshold $E_{\text{ths}}$. If $m_{0,\kappa}^\eta m_{1,\kappa}^\eta<0$, the Hamiltonian $h_\kappa^{\eta}( {\bf k})$ has a non-zero BdG D Chern number \cite{WangZhou15}, and if $\tilde{m}_{0,\kappa}^\eta<E_{\text{ths}}$, the states $\ket{\varphi_{y,\kappa}^{\eta}}$ and $\ket{\chi_{y,\kappa}^{\eta}}$ represent a chiral Majorana edge state (CMES) at each edge of the system. In this description, CMESs with different chiralities can coexist at the same edge if $m_{0,\kappa}^\eta m_{1,\kappa}^\eta<0$ and $\tilde{m}_{0,\kappa}^\eta<E_{\text{ths}}$ for several $\{\eta,\kappa\}$. This description is valid at vanishing disorder in the system. At finite disorder, CMESs of opposite chirality will hybridize, resulting in 0, 1 or several copropagating CMESs at each edge. It is worth noting that a pair of copropagating CMESs at each edge is topologically equivalent to a QAH chiral edge state \cite{QiHughes10}. If $m_{0,\kappa}^\eta m_{1,\kappa}^\eta<0$, $\tilde{m}_{0,\kappa}^\eta>E_{\text{ths}}$, and $\tilde{m}_{0,\kappa}^\eta \tilde{m}_{1,\kappa}^\eta<0$, then the states $\ket{\varphi_{x,\kappa}^{\eta}}$ and $\ket{\chi_{x,\kappa}^{\eta}}$ form a pair of MBS with negligible energy, localized at the end of the nanoribbon. If disorder were included in our description, we would either find 0 or 1 pairs of MBS in the system. 

\begin{figure}
    \includegraphics[width=\linewidth]{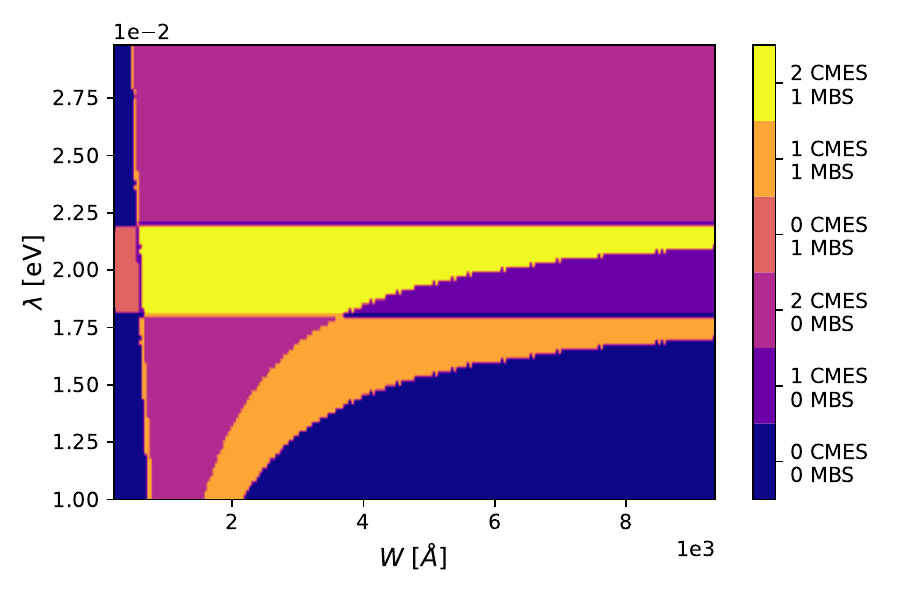}

    \caption{Number of pairs of chiral Majorana edge states (CMES) and the number of end Majorana bound states (MBS) in an MTI/SC finite nanoribbon as a function of $\lambda$ and $W$. To identify ``low-energy'' states, we retain eigenenergies below a certain threshold, here chosen to be one tenth of the superconducting pairing magnitude $|\Delta_1|$. The thickness of the slab is $d=30$ $\angstrom$ and the values of the coefficients $m_0$, $m_1$, $v_\mathrm{F}$ are taken from Table~\ref{tab1}. Moreover, $D=0$ and $\Delta_1=-\Delta_2=2 $ meV. For the results presented here, we checked that lengths $L>10W$ are sufficient.}
    \label{fig_PhaseDiagLowEStates0}
\end{figure}

As an example, let us consider a Bi$_2$Se$_3$ nanoribbon with thickness $d=30$ $\angstrom$ (see Table~\ref{tab1}) and with $m_0<0$, $m_1>0$ (the Bi$_2$Se$_3$ nanoribbon is a quantum spin hall insulator at $\lambda=0$ and $\Delta_1=\Delta_2=0$). We consider the regime where $|m_0| \gg |\Delta_1|$, which corresponds to the experimentally relevant range of superconducting pairings. We arbitrarily set $E_{\text{ths}}$ to be one tenth of the superconducting pairing magnitude $|\Delta_1|$.
In Fig.~\ref{fig_PhaseDiagLowEStates0}, we display the number of CMES and the number of MBS appearing below $E_{\text{ths}}$, as a function of $\lambda$ and $W$.

In the limit $W \rightarrow \infty$, the number of CMES is in agreement with the phase diagram of Fig.~\ref{fig41}. Moreover, for the region of Fig.~\ref{fig41} where $N=1$, and when $W$ is small enough such that there is no edge states in the $y$ direction, the system simply hosts a pair of Majorana bound states \cite{ZengYongxin18}. For intermediate range of widths, we observe a richer phase diagram. Indeed, at fixed $\lambda$, varying the width up to $\sim 10^4$ $\angstrom$, we cross different phases. Namely, we observe a region characterized by either one CMES or two CMES with the same chirality and MBS.

\begin{figure}
    \includegraphics[width=\linewidth]{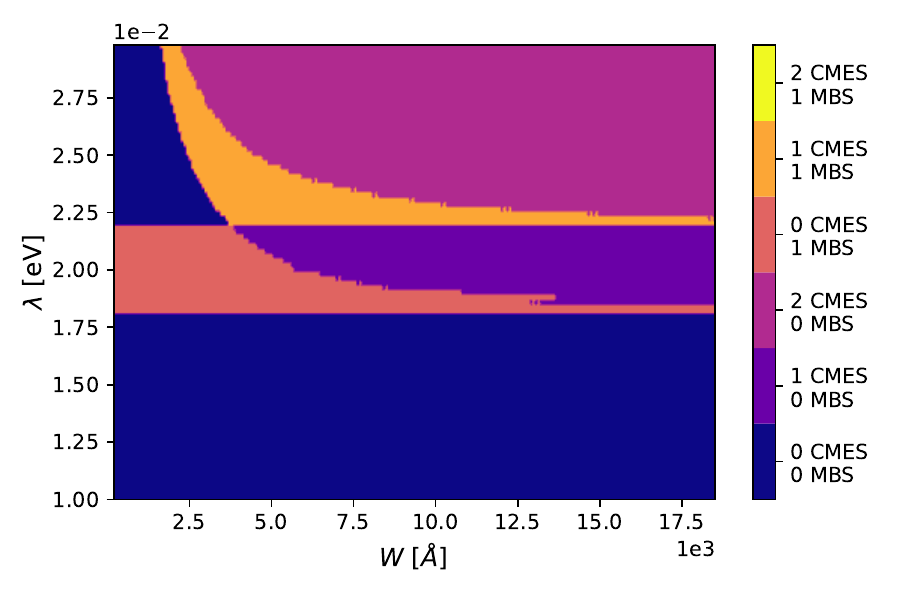}

    \caption{Number of pairs of chiral Majorana edge states (CMES) and the number of end Majorana bound states (MBS) in an MTI/SC finite nanoribbon as a function of $\lambda$ and $W$. To identify ``low-energy'' states, we retain eigenenergies below a certain threshold, here chosen to be one tenth of the superconducting pairing magnitude $|\Delta_1|$. The thickness of the slab is $d=30$ $\angstrom$ and the values of the coefficients $m_1$, $v_\mathrm{F}$ are taken from Table~\ref{tab1}. Here we considered $m_0=20$ meV, such that $m_0 m_1>0$, in order to compare with the situation $m_0 m_1<0$. Moreover, $D=0$ and $\Delta_1=-\Delta_2=2 $ meV. For the results presented here, we checked that lengths $L>10W$ are sufficient.}
    \label{fig_PhaseDiagLowEStates1}
\end{figure}

In Fig.~\ref{fig_PhaseDiagLowEStates1}, we display the number of CMESs and the number of MBSs in the case $m_0>0$, $m_1>0$ (the Bi$_2$Se$_3$ nanoribbon is a trivial insulator at $\lambda=0$ and $\Delta_1=\Delta_2=0$). For a straightforward comparison with Fig.~\ref{fig_PhaseDiagLowEStates0}, we consider $m_0$ with opposite sign and otherwise identical parameters. For wide enough nanoribbons ($W \gtrsim 2 \times 10^4 \angstrom$), we observe either 0, 1 or 2 CMESs in agreement with Fig.~\ref{fig41}. Similarly to the results shown in Fig.~\ref{fig_PhaseDiagLowEStates0}, when $W$ is small enough, we observe 0, 1, and 0 end states, respectively, for the regions of Fig.~\ref{fig41} where $N=0$, $N=1$, and $N=2$. The region of intermediate width is qualitatively different from what we observe in Fig.~\ref{fig_PhaseDiagLowEStates1}. Only in the region of Fig.~\ref{fig41} where $N=2$, we observe a phase characterized by one CMES and MBS. 

%%%%%%%%%%%%%%%%%%%%%%%%%%%%%%%%%%%%%%%%%%%%%%%%%%%%%
%%%%%%%%%%%%%%%%%%%%%%%%%%%%%%%%%%%%%%%%%%%%%%%%%%%%

%%%%%%%%%%%%%%%%%%%%%%%%%%%%%%%%%%%%%%%%%%%%%%%%%%%%%
%%%%%%%%%%%%%%%%%%%%%%%%%%%%%%%%%%%%%%%%%%%%%%%%%%%%

%%%%%%%%%%%%%%%%%%%%%%%%%%%%%%%%%%%%%%%%%%%%%%%%%%%%%
%%%%%%%%%%%%%%%%%%%%%%%%%%%%%%%%%%%%%%%%%%%%%%%%%%%%

%%%%%%%%%%%%%%%%%%%%%%%%%%%%%%%%%%%%%%%%%%%%%%%%%%%%%
%%%%%%%%%%%%%%%%%%%%%%%%%%%%%%%%%%%%%%%%%%%%%%%%%%%%

\subsection{Ribbon geometry for general parameters} \label{sec:hjsdfco1}

In the previous section, the choice of parameters $D=0$, $\Delta_1 = - \Delta_2$, and $\mu = 0$ made it possible to calculate the low-energy states for the in-plane confined geometry without further approximations. Here, we discuss the more general parameter regime $D\neq0$, $\mu \neq 0$, $\mu \ll m_0$, and $\Delta_1 \neq - \Delta_2$. 

In the following, we consider $\Delta_1=2$ meV and we write $\Delta_2=\alpha \Delta_1$. Moreover, we restrict our study to the case $-1\leq\alpha <1$, and $\mu \ll E_{3,4}$ where $E_{3,4}$ are the two highest energies associated to the Hamiltonian~\eqref{eq:ezrfgz} for the specific MTI thickness $d=30$ $\angstrom$. For this regime of parameters, we assume that each pair of energy bands around the $\Gamma$ point (see Eq.~\eqref{eq:ezrfgz}) can be described to a good approximation by a massive Dirac Hamiltonian. Therefore, close enough to the $\Gamma$ point, we express $H_{\textrm{BdG}}^{2\textrm{D}}({\bf k})$ in Eq.~\eqref{eq:ezrfgz} by $H({\bf k})=\sum_{i=1}^4H_i({\bf k})$ where
\begin{equation} \label{eq:modeleq}
    H_i({\bf k}) = v_i\left( k_y \sigma_x - k_x \sigma_y\right) + \left( M_i  - B_i k^2 \right) \sigma_z,
\end{equation}
where the Pauli matrices act in a transformed basis of states which coincides with the low-energy states of $H_{\textrm{BdG}}^{2\textrm{D}}({\bf k})$ at $\bf{k}=0$. The parameters $v_i$, $M_i$, and $B_i$ are obtained by fitting the resulting energies around the $\Gamma$ point with the corresponding energy band $E_i$ of $H_{\textrm{BdG}}^{2\textrm{D}}({\bf k})$. We further impose that the effective parameters should reduce to $m_{0,\kappa}^\eta$ and $m_{1,\kappa}^\eta$ at $\mu=0$, $D=0$, $\Delta_2 = - \Delta_1$ and that they evolve adiabatically in $(\lambda, D, \alpha, \mu)$ parameter space at constant topological invariant $N$, which is determined by the phase boundaries evolution according to Eq.~\eqref{eq:iyhwedlof}.

For sufficiently small width $W$ and $M_i B_i> 0$, the low-energy edge states associated to $H_i( k_x, -i \partial_y)$ hybridize. Similarly to Sec.~\ref{sec:hjsdfco}, a 1D Hamiltonian $\tilde{H}_i( k_x)$ can be derived from $H_i( k_x, -i \partial_y)$,
\begin{align} \label{eq:heff_kx}
\tilde{H}_{i}(k_x) =  -  \tilde{v}_i k_x \tilde{\sigma}_x + \left(\tilde{M}_i - \tilde{B}_i k_x^2 \right) \tilde{\sigma}_z,
\end{align}
where $\tilde{\sigma}_x$ and $\tilde{\sigma}_z$ are Pauli matrices acting in the $\left\{\ket{\varphi_y},\ket{\chi_y}\right\}$ basis where $\ket{\varphi_y}$ and $\ket{\chi_y}$ are the eigenstates of $H\left(k_x=0, k_y \rightarrow -i \partial_y \right)$ with the two lowest eigenenergies
$E_\pm$. The coefficients appearing in the previous equation are given by $\tilde{v}_\mathrm{F}= v  \bra{\varphi_y} \sigma_y \ket{\chi_y}$, $   2 \tilde{m}_0 = -E_+ + E_-$, $2\tilde{B} = B \left(\bra{\varphi_y} \sigma_z \ket{\varphi_y} -\bra{\chi_y} \sigma_z \ket{\chi_y} \right)$.

\subsubsection{$D \neq 0$} 
First, we consider $-1 \text{eV}\angstrom^2$ $\leq D \leq 0$ (such that $|\left(m_0-\lambda\right)D/m_1|<\left|m_0 + \lambda  \right|$ as we already consider in Sec.~\ref{sec:hjsdfco}), $\Delta_1 = - \Delta_2$ and $\mu = 0$. This case is simplified by the facts that (i) the parameter $D$ appears in the Hamiltonian with a $\gamma_z$ matrix, in contrast to the $m_1$ term proportional to $\tilde{\tau}_x \gamma_z$, and (ii) the parameter $D$ multiplies a factor $k^2$. Therefore the $D$ term does not change the value of the masses $M_i$ and has an important effect only if $D k^2 \gtrsim \Delta_1$. Since here we consider at most $D = -1$ eV$\angstrom^2$, for $k \lesssim 0.02$ $\angstrom^{-1}$ $D$ has no strong qualitative effect on the low-energy states at the system widths and lengths we consider. 

\begin{figure}
    \includegraphics[width=\linewidth]{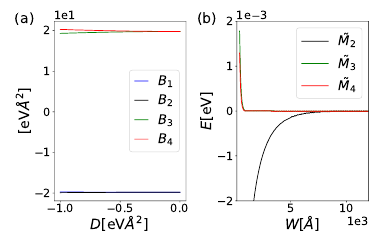}

    \caption{(a) The fitted parameters $B_i$ appearing in Eq.~\eqref{eq:modeleq}, for $\mu=0$ and $\Delta_1=-\Delta_2$. (b) The resulting low-energy spectrum for a ribbon as a function of $W$. The thickness of the slab is $30$ $\angstrom$, the values of the coefficients $m_0$, $m_1$, $v_\mathrm{F}$ are taken from Table~\ref{tab1} and $D=-1 \text{eV}\angstrom^2$. We consider $\lambda=19$ meV and $\Delta_1=-\Delta_2=2$ meV.}
    \label{figFitB}
\end{figure}

As an illustration in Fig.~\ref{figFitB}, we show the change of $B_i$ with $D$ for a specific value of $\lambda$, and we show how Fig.~\ref{fig5}(a) is modified for $D=-1$ eV$\angstrom^2$ due to the evolution of the low-energy states $\pm \Tilde{M}_i$ which appear due to the confinement along $y$ when $M_iB_i>0$.

%%%%%%%%%%%%%%%%%%%%%%%%%%%%%%%%%%%%%%%%%%%%%%%%%%%%%
%%%%%%%%%%%%%%%%%%%%%%%%%%%%%%%%%%%%%%%%%%%%%%%%%%%%

\subsubsection{$\Delta_2 \neq -\Delta_1$}
\begin{figure}
    \includegraphics[width=\linewidth]{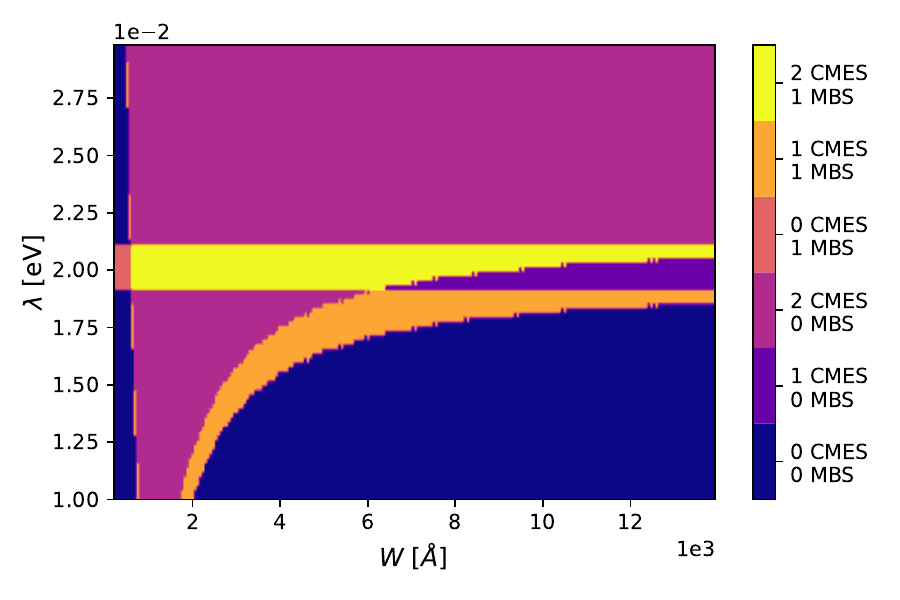}

    \caption{Number of pairs of chiral Majorana edge states (CMES) and the number of end Majorana bound states (MBS) in an MTI/SC finite nanoribbon as a function of $\lambda$ and $W$. To identify ``low-energy'' states, we retain eigenenergies below a certain threshold, here chosen to be one tenth of the superconducting pairing magnitude $|\Delta_1|$. The thickness of the slab is $d=30$ $\angstrom$ and the values of the coefficients $m_0$, $m_1$, $v_\mathrm{F}$ are taken from Table~\ref{tab1}. Moreover, $D=0$, $\Delta_1=2$ meV and $\Delta_2=0$. For the results presented here, we checked that lengths $L>10W$ are sufficient.}
    \label{fig_PhaseDiagLowEStates_alpha0}
\end{figure}

At $\Delta_2 \neq -\Delta_1$, the phase boundaries are shifted according to Eq.~\eqref{eq:iyhwedlof}. However, for each phase with a fixed topological invariant, we checked from $H({\bf k})$ that each pair of bulk bands retains the same topological character. Therefore, no strong qualitative changes happen for the phase diagram, as we show in Fig.~\ref{fig_PhaseDiagLowEStates_alpha0} for the case $\alpha=0$.

%%%%%%%%%%%%%%%%%%%%%%%%%%%%%%%%%%%%%%%%%%%%%%%%%%%%%
%%%%%%%%%%%%%%%%%%%%%%%%%%%%%%%%%%%%%%%%%%%%%%%%%%%%

\subsubsection{$\mu \neq 0$}

The case $\mu \neq 0$ is more subtle because the energies around the $\Gamma$ point can change significantly. For simplicity, we only consider values for $\lambda$ and $\mu$ corresponding to the region with topological phase $N=1$ in the infinite 2D geometry with $\text{max}(|\Delta_1|,|\Delta_2|) \lesssim \mu \ll E_{3,4}$ where $E_{3,4}$ are the two highest energies of the Hamiltonian~\eqref{eq:ezrfgz}. Moreover, we focus on the case $W >W_{\text{min}}=10^3 \angstrom$, unless otherwise stated. 
From our study in the previous section, we expect this region to be interesting since at $\mu = 0$ it shows a coexistence of two CMES and one MBS (or 1 CMES at very large $W$). How do these topological states change for $\mu \neq 0$?

Firstly, both high energies $E_{3,4}$ remain similar as in the case $\mu=0$, since here $\mu \ll E_{3,4}$. 
Therefore, above $W = 1000$ $\angstrom$, the low-energy states arising from confinement of the bulk states associated to $E_{3,4}$  
are copropagating CMES with negligible energy, as it is the case for $\mu=0$ (blue and red lines in Fig.~\ref{fig5}(a)).
Secondly, we describe the evolution of the low energy bands $E_{1,2}$ at $\mu \neq 0$ by $H({\bf k})=\sum_{i=1}^2H_i({\bf k})$ with $H_i({\bf k})$ given in Eq.~\eqref{eq:modeleq}.  

For the $N=1$ phase we consider, we know that at $W \rightarrow \infty$ the topological low-energy states correspond to only one CMES at each edge. This means that the total number of inverted bulk bands cannot change when $\mu$ changes. Taking this constraint into account, we observe from our fit that the topological characters of the first and second energy bands are exchanged at small $\mu$ with concomitant sign changes of $B_1$ and $B_2$. 
Although this has no impact on the phase diagram at $W \rightarrow \infty$, this is important for the topological properties of the low-energy states for nanoribbons of intermediate width $W$.

%%%%%%%%%%%%%%%%%%%%%%%%%%%%%%%%%%%%%%%%%%%%%%%%%%%%%
\begin{figure}
   \includegraphics[width=\linewidth]{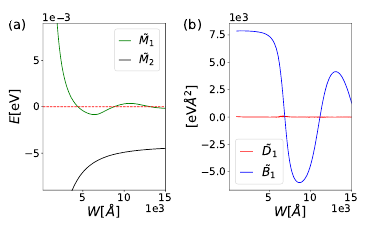}

    \caption{(a) Low-lying energies of an MTI/SC ribbon heterostructure, as a function of the width $W$ and for $\lambda=19$ meV, $\mu=3$ meV, $\Delta_1=2$ meV, and $\Delta_2=0$. (b) The coefficients $\tilde{D}_{1}$ and $\tilde{M}_{1}$, associated to the 1D bulk Hamiltonian which describe the physics of the low-energy states $\ket{\varphi_{y}}$ and $\ket{\chi_{y}}$. The thickness of the slab is $30$ $\angstrom$, the values of the coefficients $m_0$, $m_1$, $v_\mathrm{F}$ are taken from Table~\ref{tab1} and $D=-1 \text{eV}\angstrom^2$.}
    \label{fig:lowEmu}
\end{figure}
%%%%%%%%%%%%%%%%%%%%%%%%%%%%%%%%%%%%%%%%%%%%%%%%%%%%

\begin{figure}
    \includegraphics[width=1\linewidth]{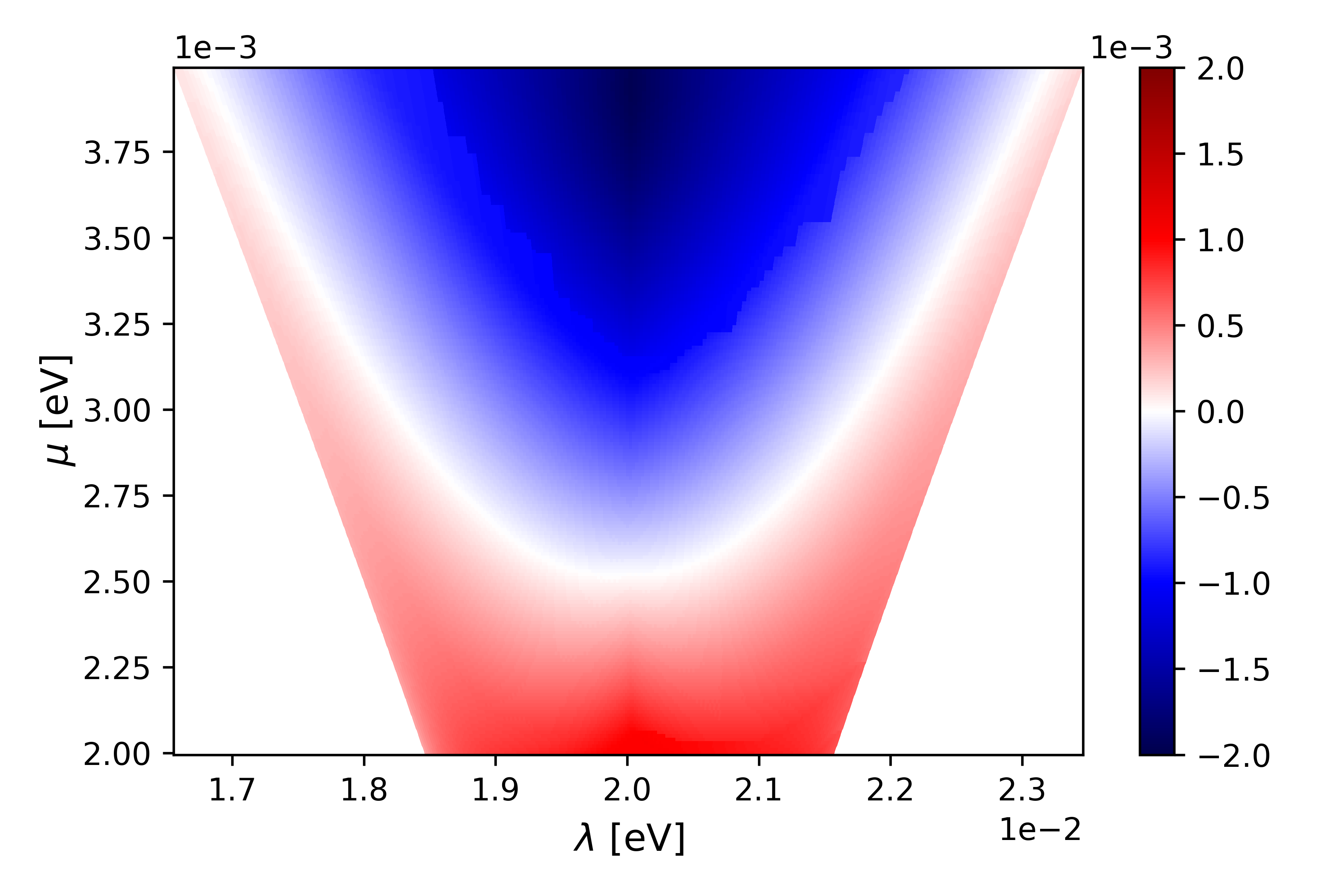}
    \caption{Gap value $\tilde{M}_1$ [see Eq.~\eqref{eq:heff_kx}] for an MTI/SC slab with thickness $d=30$ $\angstrom$ and width $W=5200$ $\angstrom$. The parameters of the 2D model, $m_0$, $m_1$, and $v_\mathrm{F}$ are taken from Table~\ref{tab1} and $D=-1 \text{eV}\angstrom^2$, and we considered $\Delta_1=2$ meV, $\Delta_2=0$. The values we consider for $\lambda$ and $\mu$ correspond to the region with topological phase $N=1$ in the infinite 2D geometry.}
    \label{fig:10}
\end{figure}

In the example of Fig.~\ref{fig:lowEmu}, we observe that $\tilde{M}_1$ takes non-negligible values for $W \lesssim 15 \times 10^3$ $\angstrom$, with an oscillating behavior as a function of $W$. This oscillating behavior is also observed as a function of $\mu$ and $\lambda$ when the width $W$ is fixed, as it is shown in Fig.~\ref{fig:10}. Moreover, the sign of $\tilde{B}_1$ and the resulting sign of $\tilde{M}_1 \tilde{B}_1$ also oscillate as function of $W$, $\mu$ and $\lambda$. For the parameters $\mu$ and $\lambda$ considered in Fig.~\ref{fig:10}, $\tilde{B}_1$ is positive, so the sign changes of $\tilde{M}_1 \tilde{B}_1$ are determined by $\tilde{M}_1 $. From our theory, this signals topological transitions as $W$, $\mu$ and $\lambda$ are varied. 

\begin{figure}
    \includegraphics[width=1\linewidth]{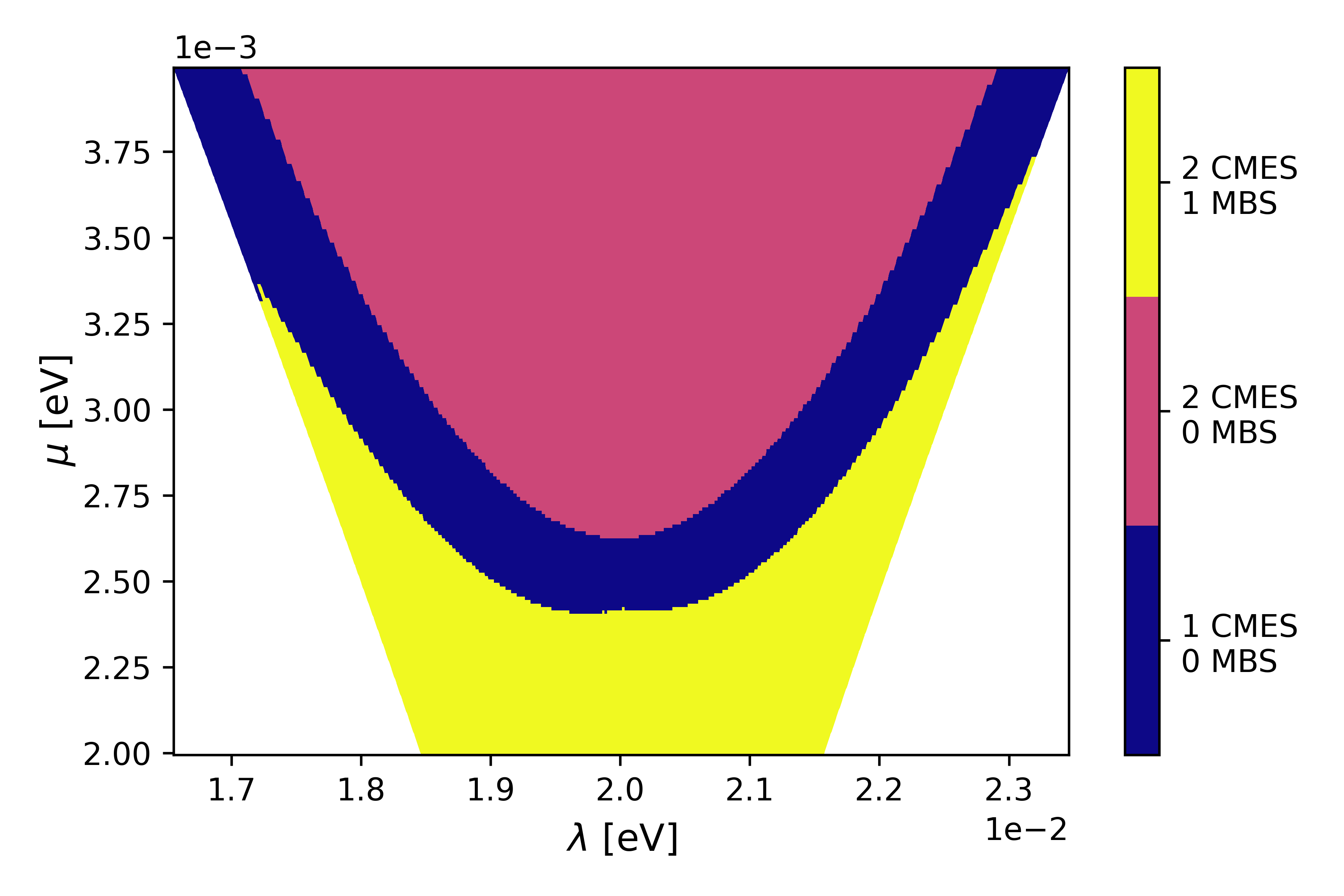}
    \caption{Number of pairs of chiral Majorana edge states (CMES) and the number of end Majorana bound states (MBS) in an MTI/SC finite nanoribbon as a function of $\lambda$ and $\mu$ and for thickness of the slab $d=30$ $\angstrom$ and width $W=5200$ $\angstrom$. To identify ``low-energy'' states, we retain eigenenergies below a certain threshold, here chosen to be one tenth of the superconducting pairing magnitude $|\Delta_1|$. The parameters of the 2D model, $m_0$, $m_1$, and $v_\mathrm{F}$ are taken from Table~\ref{tab1} and $D=-1 \text{eV}\angstrom^2$, and we considered $\Delta_1=2$ meV, $\Delta_2=0$. The values we consider for $\lambda$ and $\mu$ correspond to the region with topological phase $N=1$ in the infinite 2D geometry.}
    \label{fig:NCEM_MBS_mu}
\end{figure}

Let us also consider confinement along the $x$ direction, such that the length of the nanoribbon satisfies $L \gg W$. Here again, for concreteness, we are interested only in states with energy below a certain threshold $E_{\text{ths}}$ which we arbitrarily set to be one tenth of the superconducting pairing magnitude $|\Delta_1|$. If $\tilde{M}_{1}<E_{\text{ths}}$, the state associated to the energy $\tilde{M}_{1}$ is a CMES with negligible energy and with chirality opposite to the copropagating states associated to $E_{3,4}$. This description is valid at vanishing disorder in the system. At finite disorder, CMES of opposite chirality hybridize, resulting in only one CMES in the system. If $\tilde{M}_{1}>E_{\text{ths}}$ and $\tilde{M}_{1} \tilde{B}_{1}>0$ (see, e.g., the red phase in Fig.~\ref{fig:10}), then a pair of MBS with negligible energy, localized at the ends of the nanoribbon, appears. In this case, the low-energy states in the system are both copropagating CMES associated with $E_{3,4}$ and a pair of MBS. In Fig.~\ref{fig:NCEM_MBS_mu}, we display the number of CMES and the number of MBSs appearing at low energies as a function of $\mu$ and $\lambda$ for $W=5200\angstrom$. 

Finally, let us comment the situation in which the nanoribbon is very thin, here meaning $W\lesssim500\angstrom$, where the states arising from confinement of the bulk states associated to $E_{3,4}$ hybridize in the bulk and do not result in topological low-energy states. In this case, $\tilde{M}_{1}>E_{\text{ths}}$ and $\tilde{M}_{1} \tilde{B}_{1}>0$ (see Fig.~\ref{fig:lowEmu}), which yields a topological phase with a pair of MBS, localized at the ends of the nanoribbon, and without any CMES, reflecting the scenario proposed in Refs.~\cite{ZengYongxin18,ChenXie18}. 
Note that this topological phase can arise for $\lambda$ both greater and smaller than $|m_0|$, hence not requiring the MTI system (with $\Delta_1 = \Delta_2 = 0$) to be in the QAH phase for very large widths. While this phase only appears in a narrow window of $\lambda$ near $|m_0
|$, this region is enlarged when $\mu$ is increased, similar to the topological phase with 2 CMES and 1 MBS in Fig.~\ref{fig:NCEM_MBS_mu}.

%%%%%%%%%%%%%%%%%%%%%%%%%%%%%%%%%%%%%%%%%%%%%%%%%%%%%
%%%%%%%%%%%%%%%%%%%%%%%%%%%%%%%%%%%%%%%%%%%%%%%%%%%%

%%%%%%%%%%%%%%%%%%%%%%%%%%%%%%%%%%%%%%%%%%%%%%%%%%%%%
%%%%%%%%%%%%%%%%%%%%%%%%%%%%%%%%%%%%%%%%%%%%%%%%%%%%

\section{Conclusion}
We have studied the topological properties of finite MTI/SC heterostructures using symmetry-constrained low-energy models. We started by developing analytical models for MTI slabs with a finite thickness as well as MTI nanoribbons with finite thickness and width. We investigated the appearance of low-energy states as a function of the magnetic doping, the chemical potential and the system size. Next, we considered such finite geometries subject to superconducting pairing induced by two superconductors at the top and bottom surfaces. For very wide nanoribbons the low-energy states are the chiral edge states as predicted by the 2D bulk topological invariant. For finite width nanoribbons, we constructed and studied low-dimensional models describing the low-energy properties of our system. In a nanoribbon geometry with finite width and length, we observed regions where the low-energy states can host coexisting chiral edge states and Majorana bound states depending on the strength of the magnetic exchange term. Finally, we investigated the effect of a finite chemical potential on the topology of our system. We have studied how the bulk invariant is modified and we have built low-energy models to study the modifications in the low-energy states which appear at the boundaries of the system. When varying the magnetic doping, the chemical potential and the size of the system, we observed topological transitions between two phases which differ by the presence or absence of MBS.

\begin{acknowledgments}
This project is financially supported by the QuantERA grant MAGMA, by the National Research Fund Luxembourg under the grant INTER/QUANTERA21/16447820/MAGMA, by the German Research Foundation under grant 491798118, by MCIN/AEI/10.13039/501100011033 under project PCI2022-132927, and by the European Union NextGenerationEU/PRTR.
L.S.~acknowledges support from Grants No.~PID2020-117347GB-I00 funded by MCIN/AEI/10.13039/501100011033 and No.~PDR2020-12 funded by GOIB.
K.M.~acknowledges the financial support by the Bavarian Ministry of Economic Affairs, Regional Development and Energy under Grant No.~07 02/686 58/1/21 1/22 2/23 and by the German Federal Ministry of Education and Research (BMBF) via the Quantum Future project ‘MajoranaChips’ (Grant No. 13N15264) within the funding program Photonic Research Germany.
\end{acknowledgments}

\appendix

\section{Effective ribbon Hamiltonian via a projection onto the low-energy states} \label{app:LowEProj}

Here we determine the lowest eigenergies and the associated eigenstates of the Hamiltonian block,
\begin{equation}
    h_{0-}(-i \partial_y) = D \partial_y^2 -i  v_\mathrm{F} \sigma_x \partial_y  +  \left(m_0 - \lambda -m_1 \partial_y^2 \right)\sigma_z
\end{equation}
The energies $E$ are given by the transcendental equation
\begin{equation}
    \dfrac{ \lambda_2\left[ \xi (m_0 - \lambda) -E - \left(-D + \xi m_1 \right) \lambda_1^2 \right]}{ \lambda_1 \left[ \xi (m_0 - \lambda) -E - \left(-D + \xi m_1 \right) \lambda_2^2 \right] }
      =  \dfrac{\tanh  \lambda_2 w}{\tanh \lambda_1 w} ,
\end{equation}
with $w \equiv W/2$,
\begin{equation}
    \lambda_\alpha =  \sqrt{\dfrac{-F + (-1)^{\alpha-1} \sqrt{R}}{2\left(m_1^2+D^2\right)^2}}, \, \alpha=\{1,2\},
\end{equation}
as well as $F= v_\mathrm{F}^2 - 2 D E + 2 (m_0-\lambda) m_1$ and $R=F^2-4(D^2-m_1^2) [ E^2 - \left(m_0-\lambda\right)^2]$. The values $\xi=\pm 1$ denote, respectively, the solutions for the energy $E_+$ with eigenstate $\ket{\chi_y}$ and the energy $E_-$ with eigenstate $\ket{\varphi_y}$. We restrict the study to both low-lying energies. Their dependency with respect to the width $W$ is shown in Fig.~\ref{fig:energies_over_w_d30}.
The eigenvectors are given by
\begin{equation} \label{eq:dbkjclker3}
    \ket{\chi_y} =  N_1 \begin{pmatrix}
    -\left(-D -m_1 \right) \eta_2 f_+(y) \\
         i  v_\mathrm{F}   f_-(y) 
    \end{pmatrix},
\end{equation}
and 
\begin{equation} \label{eq:dbkjclker31}
    \ket{\varphi_y}=  N_2 \begin{pmatrix} -\left(-D -m_1 \right) \eta_1 f_-(y) \\
         i  v_\mathrm{F}   f_+(y) 
    \end{pmatrix},
\end{equation}
with 
\begin{align}
   f_+(y)&=\dfrac{\cosh \lambda_1 y}{\cosh \lambda_1 w}  - \dfrac{\cosh \lambda_2 y}{\cosh \lambda_2 w}, \notag \\ f_-(y)&=\dfrac{ \sinh \lambda_1 y}{\sinh \lambda_1 w} - \dfrac{ \sinh \lambda_2 y}{\sinh \lambda_2 w}, \notag \\
   \eta_2 &= \dfrac{ \lambda_1^2 -\lambda_2^2 }{\lambda_1 \tanh \lambda_1 w -\lambda_2 \tanh \lambda_{2} w}, \notag \\ \eta_1 &= \dfrac{\left[ \lambda_1^2 -\lambda_2^2  \right]}{\lambda_1 \coth{\lambda_1 w } -\lambda_2 \coth{\lambda_2 w }},
\end{align}
and normalization constants $N_1$ and $N_2$.

Next we project $h_{1-}(k_x)$,
\begin{equation}
    h_{1-}(k_x) = -\mu - D k_x^2 +  v_\mathrm{F}  k_x \sigma_y   +m_1 k_x^2 \sigma_z,
\end{equation}
on the low-energy eigenstates of $h_{0-}(-i \partial_y)$ and we obtain an effective Hamiltonian for the system initially described by $h_{-}^*$, valid around $k_x=0$. Here we consider only the two low-energy eigenstates of $h_{0-}(-i \partial_y)$ and we assume that these eigenstates are well separated in energy from the other eigenstates. This assumption is valid for the mass inverted regime $\left(m_0- \lambda \right)m_1 <0$ we are considering here.
Then we find the Hamiltonian given by the Eq.~\eqref{eq:HMTIprojectedW}.

\section{Transforming the Hamiltonian into a sum of Dirac terms in spin space} \label{app:DiagBlocksH}

The matrix $\sigma_z \tilde{\tau}_x$ commutes with the Hamiltonian at $\mu=0$ and $D=0$ and $\Delta_1 =- \Delta_2$,
\begin{align}
\begin{split}
    H_{\textrm{BdG}}^{2\textrm{D}}({\bf k}) 
=  
    & v_\mathrm{F} k_y \sigma_x \tilde{\tau}_z  + \big[ - v_\mathrm{F} k_x \sigma_y \tilde{\tau}_z + \lambda \sigma_z \\ 
&+ 
     m({\bf k})  \tilde{\tau}_x  \big] \gamma_z - 
    \Delta_1 \tilde{\tau}_z \sigma_y \gamma_y.
\end{split}
\end{align}
Therefore it is possible to diagonalize $H_{\textrm{BdG}}^{2\textrm{D}}({\bf k})$ and $\sigma_z \tilde{\tau}_x$ using a common basis transformation defined via a unitary matrix $U$ which diagonalizes $\sigma_z \tilde{\tau}_x$. For $U = ( \tilde{\tau}_z + \tilde{\tau}_x )/\sqrt{2}$ one finds $U\sigma_z \tilde{\tau}_x U^\dagger= \sigma_z \tilde{\tau}_z$. This diagonal matrix has only two different eigenvalues $\pm 1$. It is convenient to define $P$ as the projector on the eigenspace with eigenvalue $1$ whereas $\overline{P}=1-P$ projects on the eigenspace with eigenvalue $-1$. Then we have, 
\begin{align}
    H_{\textrm{BdG}} &= \sum_{\bf k} \psi_{\bf k}^\dagger U^\dagger P U H_{\textrm{BdG}}({\bf k}) U^\dagger P U \psi_{\bf k} \notag \\ &+ \sum_{\bf k} \psi_{\bf k}^\dagger U^\dagger \overline{P} U H_{\textrm{BdG}}({\bf k}) U^\dagger \overline{P} U \psi_{\bf k}. 
\end{align}
Next, we have
\begin{align}
    U H_{\textrm{BdG}}({\bf k}) U^\dagger &= v_\mathrm{F} k_y \sigma_x \tilde{\tau}_x - |\Delta| \tilde{\tau}_x \sigma_y \gamma_y \notag \\ &+ \left[ - v_\mathrm{F} k_x \sigma_y \tilde{\tau}_x + \lambda \sigma_z  +  m({\bf k})  \tilde{\tau}_z \right] \gamma_z.
\end{align}
Choosing $P =\left( \sigma_z \tilde{\tau}_z + 1 \right)/ 2$ and using $\tilde{\tau}_z P=\sigma_z P$ and $\tilde{\tau}_z \overline{P}=-\sigma_z \overline{P}$, we write 
\begin{align}
    H_{\textrm{BdG}}&= \sum_{\bf k} \psi_{\bf k}^\dagger U^\dagger P H_{\textrm{BdG}}^P({\bf k})  P U \psi_{\bf k} \notag \\ &+ \sum_{\bf k} \psi_{\bf k}^\dagger U^\dagger \overline{P} H_{\textrm{BdG}}^{\overline{P}}({\bf k}) \overline{P} U \psi_{\bf k},
\end{align}
with 
\begin{align}
H_{\textrm{BdG}}^{P,\overline{P}}({\bf k}) &=  A_2 k_y \sigma_x \tilde{\tau}_x  -A_2 k_x \sigma_y \tilde{\tau}_x \gamma_z \notag \\ & + \left[ \lambda \pm m({\bf k})  \right] \sigma_z  \gamma_z - |\Delta| \tilde{\tau}_x \sigma_y \gamma_y,
\end{align}

Now we see that applying a basis transformation which diagonalizes $\tilde{\tau}_x$ makes the Hamiltonian diagonal in \{top,bottom\} space. Therefore we perform another unitary transformation using $U$,
\begin{align}
    H_{\textrm{BdG}}= &\sum_{\bf k} \psi_{\bf k}^{P \dagger} U H_{\textrm{BdG}}^P({\bf k}) U^\dagger \psi_{\bf k}^P \notag \\ &+ \sum_{\bf k} \psi_{\bf k}^{\overline{P} \dagger} U H_{\textrm{BdG}}^{\overline{P}}({\bf k}) U^\dagger \psi_{\bf k}^{\overline{P}},
\end{align}
with $\psi_{\bf k}^P =U P U \psi_{\bf k}$ and $\psi_{\bf k}^{\overline{P}} =U \overline{P} U \psi_{\bf k}$ and we have
\begin{align}
   U H_{\textrm{BdG}}^{P,\overline{P}}({\bf k}) U^\dagger &=  A_2 k_y \sigma_x \tilde{\tau}_z  -A_2 k_x \sigma_y \tilde{\tau}_z \gamma_z \notag \\ &+ \left[ \lambda  \pm m({\bf k})  \right] \sigma_z  \gamma_z - |\Delta|  \sigma_y \tilde{\tau}_z \gamma_y, 
\end{align}
Now we almost have a Hamiltonian which is diagonal in \{top,bottom\} space and in Nambu space. Only the last term in the two previous equations remains off-diagonal. Another unitary transformation, which we denote by $U_1 = \left(\sigma_x+i\gamma_y \right)/\sqrt{2}$, makes the total Hamiltonian diagonal in \{top,bottom\} space and in Nambu space, 
\begin{equation}
    H_{\textrm{BdG}}= H_{\textrm{BdG},1} + H_{\textrm{BdG},2},
\end{equation}
with $ H_{\textrm{BdG},1} = \sum_{\bf k} \psi_{\bf k}^{P \dagger} U_1^\dagger H_{1}^P({\bf k}) U_1 \psi_{\bf k}^P $, $H_{\textrm{BdG},2} = \sum_{\bf k} \psi_{\bf k}^{\overline{P} \dagger} U_1^\dagger H_{1}^{\overline{P}} U_1 \psi_{\bf k}^{\overline{P}}$
and $H_{1}^P({\bf k}) = U_1 U H_{\textrm{BdG}}^P({\bf k}) U^\dagger U_1^\dagger$, $H_{1}^{\overline{P}}=U_1 U H_{\textrm{BdG}}^{\overline{P}}({\bf k}) U^\dagger U_1^\dagger$. Then, we indeed have
\begin{align}
   H_{1}^{P,\overline{P}}({\bf k}) &=  A_2 k_y \sigma_x \tilde{\tau}_z  +A_2 k_x \sigma_y \tilde{\tau}_z \gamma_z \notag \\ &- \left[ \lambda  \pm  m({\bf k})  \right] \sigma_z  \gamma_z + |\Delta|  \sigma_z \tilde{\tau}_z , 
\end{align}
The Hamiltonian is now diagonal in \{top,bottom\} space and in Nambu space. We now project it over the eigensubspace of $\gamma_z$ and of $\tilde{\tau}_z$.

First we define $P_{\gamma_z} = \left(\gamma_z + 1 \right)/ 2$ as the projector on the eigenspace with eigenvalue $+1$ of $\gamma_z$ and $\overline{P}_{\gamma_z} = 1 - P_{\gamma_z}$ projects on the eigenspace with eigenvalue $-1$ of $\gamma_z$. Then we obtain
\begin{align}
 H_{\textrm{BdG},1}&=\sum_{\bf k} \psi_{\bf k}^{0,P\dagger} H_{1}^{0,P}({\bf k}) \psi_{\bf k}^{0,P} \notag \\ &+\sum_{\bf k} \psi_{\bf k}^{1,P\dagger} H_{1}^{1,P}({\bf k}) \psi_{\bf k}^{1,P},
\end{align}
with $\psi_{\bf k}^{0,P} = P_{\gamma_z} U_1 \psi_{\bf k}^P$, and $\psi_{\bf k}^{1,P} = \overline{P}_{\gamma_z} U_1 \psi_{\bf k}^P$, and
\begin{align}
   H_{1}^{(0,1),P}({\bf k}) &=  A_2 k_y \sigma_x  \tilde{\tau}_z +A_2 k_x \sigma_y  \tilde{\tau}_z \notag \\ &\mp \left[ \lambda  +  m({\bf k})  \right] \sigma_z + |\Delta|  \sigma_z  \tilde{\tau}_z.
\end{align}
Moreover we notice that the unitary transformation $\sigma_y \tilde{\tau}_x $ performed in the eigenspace with eigenvalue $+1$ of $\gamma_z$ gives
\begin{equation}
   \sigma_y \tilde{\tau}_x H_{1}^{0,P}({\bf k}) \sigma_y \tilde{\tau}_x = H_{1}^{1,P}({\bf k})  ,
\end{equation}
and because
\begin{align}
    \psi_{\bf k}^{0,P\dagger} \sigma_y \tilde{\tau}_x H_{1}^{1,P}({\bf k}) \psi_{\bf k}^{1,P} = 0,
\end{align}
we obtain
\begin{align}
 H_{\textrm{BdG},1}= \sum_{\bf k}  \tilde{\psi}_{\bf k}^{P\dagger} H_{1}^{1,P}({\bf k}) \tilde{\psi}_{\bf k}^{P	} .
\end{align}
with $\tilde{\psi}_{\bf k}^{P\dagger} = ( \psi_{\bf k}^{0,P\dagger} \sigma_y \tilde{\tau}_x+\psi_{\bf k}^{1,P\dagger})$.
Similarly, defining $\psi_{\bf k}^{0,{\overline{P}}} = P_{\gamma_z} U_1 \psi_{\bf k}^{\overline{P}}$, $\psi_{\bf k}^{1,{\overline{P}}} = \overline{P}_{\gamma_z} U_1 \psi_{\bf k}^{\overline{P}}$ and 
\begin{equation}
   H_{1}^{1,{\overline{P}}}({\bf k}) = A_2 k_y \sigma_x  \tilde{\tau}_z -A_2 k_x \sigma_y  \tilde{\tau}_z + \left[ \lambda  -  m({\bf k})  \right] \sigma_z + |\Delta|  \sigma_z  \tilde{\tau}_z ,
\end{equation}
we obtain
\begin{align}
 H_{\textrm{BdG},2}= &\sum_{\bf k} \tilde{\psi}_{\bf k}^{{\overline{P}}\dagger} H_{1}^{1,{\overline{P}}}({\bf k}) \tilde{\psi}_{\bf k}^{{\overline{P}}},
\end{align}
with $\tilde{\psi}_{\bf k}^{{\overline{P}}\dagger} = ( \psi_{\bf k}^{0,{\overline{P}}\dagger} \sigma_y \tilde{\tau}_x+\psi_{\bf k}^{1,{\overline{P}}\dagger})$. Hence, $H_{\textrm{BdG}}$ now reads
\begin{align}
    H_{\textrm{BdG}} &= \sum_{\bf k} \tilde{\psi}_{\bf k}^{P\dagger} H_{1}^{1,P}({\bf k}) \tilde{\psi}_{\bf k}^{P} \notag \\ &+ \sum_{\bf k} \tilde{\psi}_{\bf k}^{{\overline{P}}\dagger} H_{1}^{1,{\overline{P}}}({\bf k}) \tilde{\psi}_{\bf k}^{{\overline{P}}},
\end{align}
Now we project this Hamiltonian over the \{top,bottom\} space. We define $P_{\tilde{\tau}_z} = ( \tilde{\tau}_z + 1 )/ 2$ as the projector on the eigenspace with eigenvalue $+1$ of $\tilde{\tau}_z$ and $\overline{P}_{\tilde{\tau}_z} = 1 - P_{\tilde{\tau}_z}$. Then we have
 \begin{align}
 &\sum_{\bf k} \tilde{\psi}_{\bf k}^{P\dagger} H_{1}^{1,P}({\bf k}) \tilde{\psi}_{\bf k}^{P}=\notag \\ &\sum_{\bf k} \psi_{\bf k}^{0,1,P \dagger} H_{1}^{0,1,P}({\bf k}) \psi_{\bf k}^{0,1,P} + \sum_{\bf k} \psi_{\bf k}^{1,1,P \dagger} H_{1}^{1,1,P}({\bf k}) \psi_{\bf k}^{1,1,P},
\end{align}
with $\psi_{\bf k}^{0,1,P} =  P_{\tilde{\tau}_z} \tilde{\psi}_{\bf k}^{P}$
and $\psi_{\bf k}^{1,1,P} =  \overline{P}_{\tilde{\tau}_z} \tilde{\psi}_{\bf k}^{P}$ and
\begin{align}
    &H_{1}^{0,1,P}({\bf k}) = A_2 k_y \sigma_x  -A_2 k_x \sigma_y  + \left[ \lambda  +  m({\bf k})  \right] \sigma_z + |\Delta|  \sigma_z , \notag \\
    &H_{1}^{1,1,P}({\bf k}) = -A_2 k_y \sigma_x  +A_2 k_x \sigma_y  + \left[ \lambda  +  m({\bf k})  \right] \sigma_z - |\Delta|  \sigma_z ,
\end{align}
Let us define
\begin{equation}
    h_\pm^{P }({\bf k}s)=A_2 k_y \sigma_x  -A_2 k_x \sigma_y  + \left[ \lambda  +  m({\bf k})   \pm |\Delta| \right] \sigma_z
\end{equation}
Then we have $ H_{1}^{0,1,P}({\bf k}) = h_+^{P }({\bf k})$ and $\sigma_y  H_{1}^{1,1,P}({\bf k}) \sigma_y = - h_-^{P *}(-{\bf k})$ so we conclude that 
 \begin{align}
 \sum_{\bf k} \psi_{\bf k}^{P \dagger} U_1^\dagger H_{1}^P({\bf k}) U_1 \psi_{\bf k}^P=&\sum_{\bf k} \tilde{\psi}_{\bf k}^{P \dagger} \tilde{H}_{1}^{P}({\bf k}) \tilde{\psi}_{\bf k}^{P},
\end{align}
with $\tilde{\psi}_{\bf k}^{P} = \left(\psi_{\bf k}^{0,1,P}, \sigma_y \psi_{\bf k}^{1,1,P} \right)^T$ and
\begin{equation}
    \tilde{H}_{1}^{P}({\bf k}) = \begin{pmatrix}
       h_+^{P}({\bf k}) &0 \\
        0&- h_-^{P *}(-{\bf k})
    \end{pmatrix}
\end{equation}
The same transformations can be applied to the other part of the Hamiltonian and we find
 \begin{align}
 \tilde{\psi}_{\bf k}^{{\overline{P}}\dagger} H_{1}^{1,{\overline{P}}}({\bf k}) \tilde{\psi}_{\bf k}^{{\overline{P}}}=\sum_{\bf k} \tilde{\psi}_{\bf k}^{{\overline{P}} \dagger} \tilde{H}_{1}^{\overline{P}}({\bf k}) \tilde{\psi}_{\bf k}^{\overline{P}},
\end{align}
with $\tilde{\psi}_{\bf k}^{\overline{P}} = \left(\psi_{\bf k}^{0,1,{\overline{P}}}, \sigma_y \psi_{\bf k}^{1,1,{\overline{P}}} \right)^T$ and
\begin{equation}
    \tilde{H}_{1}^{\overline{P}}({\bf k}) = \begin{pmatrix}
       h_+^{\overline{P}}({\bf k}) &0 \\
        0&- h_-^{\overline{P} *}(-{\bf k})
    \end{pmatrix},
\end{equation}
with $\psi_{\bf k}^{0,1,{\overline{P}}} =  P_{\tilde{\tau}_z} (\sigma_y \tilde{\tau}_x P_{\gamma_z} U_1 \psi_{\bf k}^{\overline{P}} +  \overline{P}_{\gamma_z} U_1 \psi_{\bf k}^{\overline{P}} )$, $\psi_{\bf k}^{1,1,{\overline{P}}} =  \overline{P}_{\tilde{\tau}_z} (\sigma_y \tilde{\tau}_x P_{\gamma_z} U_1 \psi_{\bf k}^{\overline{P}} +  \overline{P}_{\gamma_z} U_1 \psi_{\bf k}^{\overline{P}} )$ and
\begin{equation}
    h_\pm^{\overline{P}}({\bf k})=A_2 k_y \sigma_x  -A_2 k_x \sigma_y  + \left[ \lambda  -  m({\bf k})   \pm |\Delta| \right] \sigma_z,
\end{equation}
To sum up we have
\begin{align}
    H_{\textrm{BdG}}= \sum_{\bf k} \tilde{\psi}_{\bf k}^{P \dagger} \tilde{H}_{1}^{P}({\bf k}) \tilde{\psi}_{\bf k}^{P} +  \sum_{\bf k} \tilde{\psi}_{\bf k}^{{\overline{P}} \dagger} \tilde{H}_{1}^{\overline{P}}({\bf k}) \tilde{\psi}_{\bf k}^{\overline{P}},
\end{align}
with 
\begin{align} 
    &\tilde{H}_{1}^{P}({\bf k}) = \begin{pmatrix}
       h_+^{P}({\bf k}) &0 \\
        0&- h_-^{P *}(-{\bf k})
    \end{pmatrix}, \notag \\ &\tilde{H}_{1}^{\overline{P}}({\bf k}) = \begin{pmatrix}
       h_+^{\overline{P}}({\bf k}) &0 \\
        0&- h_-^{\overline{P} *}(-{\bf k})
    \end{pmatrix},
\end{align}
and
\begin{align}
    &h_\pm^{P}({\bf k})=A_2 k_y \sigma_x  -A_2 k_x \sigma_y  + \left[ \lambda  +  m({\bf k})   \pm |\Delta| \right] \sigma_z, \notag \\
    &h_\pm^{\overline{P}}({\bf k})=A_2 k_y \sigma_x  -A_2 k_x \sigma_y  + \left[ \lambda  -  m({\bf k})   \pm |\Delta| \right] \sigma_z.
\end{align}
We denote the energies of $h_+^{P}({\bf k})$, $- h_-^{P *}(-{\bf k})$, $h_+^{\overline{P}}({\bf k})$ and $- h_-^{\overline{P} *}(-{\bf k})$ by $E_1$, $E_2$, $E_3$, and $E_4$, respectively. We have
\begin{align}
    E_1^\pm &= \pm \sqrt{A_2^2 k_\perp^2 + \left( \lambda  +  m({\bf k}) + |\Delta|  \right)^2}, \notag \\
    E_2^\pm &= \pm \sqrt{A_2^2 k_\perp^2 + \left( \lambda  +  m({\bf k}) - |\Delta|  \right)^2}, \notag \\
    E_3^\pm &= \pm \sqrt{A_2^2 k_\perp^2 + \left( \lambda  -  m({\bf k}) + |\Delta|  \right)^2}, \notag \\
    E_4^\pm &= \pm \sqrt{A_2^2 k_\perp^2 + \left( \lambda  -  m({\bf k}) - |\Delta|  \right)^2}.
\end{align}
Note that when the system has time-reversal symmetry at $\lambda = 0$, we have $E_1^\pm = E_4^\pm$ and $E_2^\pm = E_3^\pm$. Generally speaking, in presence of time reversal symmetry, Kramers' theorem tells us that every (spin-$1/2$) Bloch state is degenerate with its time-reversal conjugate, i.e., a state $\ket{\psi_a ({\bf k})}$ with energy $E_a({\bf k})$ and a state $\ket{\psi_b(-{\bf k})} = \Theta \ket{\psi_a({\bf k})}$, where $\Theta$ is the time-reversal operator, with energy $E_b(-{\bf k})$ have the same energies, $E_a({\bf k}) = E_b(-{\bf k})$. Additionally, we have $E_i^\pm({\bf k}) = E_i^\pm(-{\bf k})$ ($i\in\{1,2,3,4\}$) due to the additional presence of inversion symmetry, so that $E_a({\bf k}) = E_b({\bf k})$, meaning that the energy bands which come in Kramers pairs $E_a({\bf k}) = E_b(-{\bf k})$ are not only degenerate at the time-reversal invariant points ${\bf k}=-{\bf k}$ but at each ${\bf k}$ point. From $E_1^\pm = E_4^\pm$ and $E_2^\pm = E_3^\pm$, we find that $E_1^-$ and $ E_4^-$ form a Kramers pair of energy bands and the same is true for $E_1^+$ and $ E_4^+$, $E_2^-$ and $ E_3^-$ and $E_2^+$ and $ E_3^+$.

\bibliography{ref}

\end{document}